\documentclass[12pt,a4paper]{article}
\usepackage{cite,amsmath,amssymb,bbm,graphicx}
\usepackage{mathrsfs}	
\usepackage{longtable,setspace,color}

\newcommand{\be}{\begin{equation}}  
\newcommand{\ee}{\end{equation}}

\newcommand\Tspace{\rule{0pt}{2.6ex}}       % Top strut
\usepackage{color}

\begin{document}

\thispagestyle{empty}

\vspace*{-2cm}
\begin{flushright}
\end{flushright}
\vspace*{1.2cm}

\begin{center}

{\LARGE\bf $\boldsymbol{R}$ symmetries and a heterotic MSSM}\\[12mm]

{\large Rolf Kappl, Hans Peter Nilles, 
Matthias Schmitz} 
\\[6mm]

{\it Bethe Center for Theoretical Physics\\
and\\
Physikalisches Institut der Universit\"at Bonn\\
Nussallee 12, 53115 Bonn, Germany
}

\vspace*{12mm}

\begin{abstract}
We employ powerful techniques based on Hilbert- and Gr\"obner bases
to analyze particle physics models derived from string theory.
Individual models are shown to have a huge landscape of vacua
that differ in their phenomenological properties. We explore
the (discrete) symmetries of these vacua, the new $R$ symmetry selection rules and their consequences for moduli stabilization. 
\end{abstract}

\end{center}

\clearpage

\section{Introduction}

Recently new results on $R$ symmetries in heterotic orbifolds have been obtained \cite{Nilles:2013lda,Bizet:2013wha}. We discuss the implications of these results on phenomenologically appealing models like the ones obtained in \cite{Lebedev:2006kn,Lebedev:2007hv,Lebedev:2008un}. 
Our purpose is twofold. On the one hand we show that with the new $R$ symmetries successful models remain. On the other hand we employ techniques developed in \cite{Kappl:2011vi} to study the vacuum configuration of a given model. These techniques enable us to determine the structure of the superpotential in form of building monomials. The advantage to previous attempts is, that one can immediately see which standard model singlets induce which couplings. We further show that this approach is also useful for general particle physics models with continuous or discrete symmetries. All of these symmetries lead to Diophantine equations whose solutions are given by so-called Hilbert basis elements. After the determination of the superpotential we search for suitable SUSY preserving minima of our model. We find that some of the standard model singlets will get stabilized at a non-trivial value. In contrast to \cite{Kappl:2010yu} many singlets will have flat directions. This can be understood in terms of remnant symmetries.

The paper is outlined as follows. In section \ref{sec:hilbert} we show how symmetries result in Diophantine equations and how these can be solved to determine the structure of the underlying physical model. In section \ref{sec:r} we briefly review new results for \(R\) symmetries in the heterotic orbifold context obtained in \cite{Nilles:2013lda,Bizet:2013wha} and comment on their consequences. We finally apply the developed techniques to a phenomenologically appealing model in section \ref{sec:pheno} and summarize our results.

\section{Allowed couplings and Hilbert bases}
\label{sec:hilbert}

It has been outlined in \cite{Kappl:2011vi} how to use Hilbert bases to compute a basis of all allowed monomials in the superpotential. In this section we will use this technique to derive which couplings are allowed by the \(R\) symmetries and the other well known string selection rules. We will explain the approach by an example from flavor model building \cite{Antusch:2014poa}
and apply the technique to our concrete string model.

\subsection{Symmetries and Diophantine equations}
\label{sec:chen}

Given some continuous or discrete symmetries one is usually interested in the question which couplings are allowed by the given symmetries. Let us assume a theory with a U(1) symmetry and fields \(\phi_i\) with charges \(q_i\) under the symmetry. An example for an allowed monomial would be
\begin{equation}
W\supset \phi_1\phi_2\phi_3^2\qquad\Leftrightarrow\qquad q_1+q_2+2q_3=0\,.
\end{equation} 
We can generalize this to
\begin{equation}
W\supset \phi_1^{n_1}\ldots\phi_M^{n_M}\qquad \Leftrightarrow\qquad q^T\cdot n=0\,,
\end{equation}
where \(q^T=(q_1,\ldots,q_M)\) and \(n^T=(n_1,\ldots,n_M)\in \mathbbm{N}_0^M\). Because of the restriction \(n_i\in\mathbbm{N}_0\) this is called a Diophantine equation. That means every Abelian gauge symmetry leads naturally to a Diophantine equation. It is not possible to give an analytical basis for all solution vectors \(n\) because the natural numbers only form a monoid. Nevertheless it is possible to algorithmically determine a basis. We can write
\begin{equation}
n=\sum_i\alpha_i x_i,\qquad x_i\in\mathcal{H}, \qquad \alpha_i\in \mathbbm{N}_0\,.
\end{equation}
Note that the total number of Hilbert basis elements \(x_i\) cannot be determined analytically.
The basis spanned by the elements \(x_i\) is called Hilbert basis and several freely available programs \cite{normaliz:2014,4ti2} exist to determine it. This was shown and outlined in more detail in \cite{Kappl:2011vi}. Let us consider an example from flavor model building, namely the model discussed in \cite{Antusch:2014poa}. Some of its fields and their respective charges under a large number of discrete symmetries are given in table \ref{tab:antusch}.
\begin{table}[htb]
\centering
\begin{tabular}{ccccccccc}
&$\mathbbm{Z}_2$&$\mathbbm{Z}_4$&$\mathbbm{Z}_4$&$\mathbbm{Z}_4$&
$\mathbbm{Z}_7$&$\mathbbm{Z}_7$&$\mathbbm{Z}_9$&$\mathbbm{Z}_2$\\
\hline
$H_5$&0&0&0&0&0&0&0&0\\
$\mathcal{T}_1$&0&0&3&0&6&0&0&1\\
$\mathcal{T}_2$&0&0&0&0&6&0&0&1\\
$\mathcal{T}_3$&0&0&0&0&0&0&0&1\\
$\theta_1$&0&0&0&0&1&0&0&0\\
$\theta_2$&0&0&1&0&0&0&0&0\\
$\theta_3$&0&2&0&0&0&6&5&0\\
$\theta_4$&0&0&0&0&0&0&5&0\\
\end{tabular}
\caption{Charges of the fields considered in the flavor model discussed in \cite{Antusch:2014poa}.}
\label{tab:antusch}
\end{table}
We are interested in Yukawa couplings involving two fields \(\mathcal{T}_i\) and a neutral Higgs field
\(H_5\)\,.
We can compute the Hilbert basis \(\mathcal{H}\) for these couplings with the help of \texttt{normaliz} \cite{normaliz:2014, 2012arXiv1206.1916B} and find the following elements in this model
\begin{equation}
\begin{split}
W&=H_5\left(\mathcal{T}_1\mathcal{T}_1(\theta_1^2\theta_2^2 + \theta_1^2\theta_2^6 + \theta_1^9\theta_2^2 + \theta_1^2\theta_2^2\theta_4^9 + \ldots) 
+\mathcal{T}_1\mathcal{T}_2(\theta_1^2\theta_2+\ldots)
+\mathcal{T}_2\mathcal{T}_2(\theta_1^2+\ldots)\right.\\
&\quad\left.+\mathcal{T}_2\mathcal{T}_3(\theta_1+\ldots)
+\mathcal{T}_3\mathcal{T}_3(1+\ldots)
\right)\,.
\end{split}
\end{equation} 
This result is in agreement with the superpotential given in \cite{Antusch:2014poa} and shows the applications of our approach. The dots denote monomials which consist of higher order couplings in the flavon fields \(\theta_i\). Note that while we considered only Abelian symmetries here, it would also be interesting to consider Hilbert bases for non-Abelian discrete symmetries like the ones discussed for example in \cite{Chen:2012st}.

\subsection{Diophantine equations and $\boldsymbol{R}$ symmetries}
\label{sec:rsym}

\(R\) symmetries result in inhomogeneous Diophantine equations. In this section we will already have a concrete string model in 
mind (benchmark model 1 from \cite{Lebedev:2007hv}). Nonetheless, the results are applicable to any model with $R$ symmetries. Our approach also deviates slightly from \cite{Kappl:2011vi}. Let us assume that our theory gives rise to the $R$ symmetry\footnote{As outlined, for example, in \cite{Kappl:2010yu} the charges under these symmetries are not integers, which means that a redefinition seems to be natural. Nevertheless we keep this form for easier comparison with the literature.} \linebreak\(\mathbbm{Z}^R_6\times \mathbbm{Z}^R_3\times \mathbbm{Z}^R_2\). This is for example the case for models in the mini-landscape and has been reconsidered recently in \cite{Nilles:2013lda,Bizet:2013wha} which is one motivation of our work. Let us denote the charges under these symmetries by \(R^T=(R^1,R^2,R^3)\). Thus the $R$ symmetry of the model will result in the constraint
\begin{equation}
\sum_i R_i^T=(-1,-1,-1)\mod (6, 3, 2)\,,
\end{equation}
where \(R_i\) denotes the \(R\) charge of a field \(\phi_i\) and we assume that the superpotential transforms with \(R\) charge one.
Let us illustrate our approach with an example. If we want to know which singlets \(\tilde{s}_i\) induce the Yukawa term
\begin{equation}
W\supset \bar{\phi}_1q_1\bar{u}_1 f(\tilde{s}_i)\,,
\end{equation}
we have to solve the inhomogeneous system of Diophantine equations
\begin{equation}
\hat{R}\cdot n=\begin{pmatrix}
-1\\
-1\\
-1
\end{pmatrix}
\mod
\begin{pmatrix} 
6\\
3\\
2
\end{pmatrix}
\,, \qquad n_i\in \mathbbm{N}_0 \quad \forall i\,,
\end{equation}
with
\begin{equation}
\hat{R}=\begin{pmatrix}
R_1&\ldots& R_{M+3}
\end{pmatrix},\qquad R_i^T=(R_i^1,R_i^2,R_i^3)\,,
\end{equation}
where \(M\) is the number of different standard model singlets in our model. During this discussion we use the notation from \cite{Lebedev:2007hv} and call the up type Higgs \(\bar{\phi}_1\) and the down type Higgs \(\phi_1\). 
Further, as in section \ref{sec:chen}, \(n_i\) denotes the exponent of a given field
\begin{equation}
W\supset \bar{\phi}_1^{n_1}q_1^{n_2}\bar{u}_1^{n_3} (\tilde{s}_1)^{n_4}(\tilde{s}_2)^{n_5}
\ldots (\tilde{s}_M)^{n_{M+3}}\,.
\end{equation} 
Only solutions with \(n_1=n_2=n_3=1\) are physical, therefore we can simplify this equation system by substituting
\begin{equation}
\tilde{R}=R_1+R_2+R_3+(1,1,1)^T\,.
\end{equation}
This results in a homogeneous system of equations with less variables
\begin{equation}
\label{equa:hom}
\begin{pmatrix}\tilde{R}& R_4&\ldots&R_{M+3}
\end{pmatrix}\cdot 
\begin{pmatrix}
n_1\\
n_4\\
\vdots\\
n_{M+3}
\end{pmatrix}
=
\begin{pmatrix}
0\\
0\\
0
\end{pmatrix}
\mod 
\begin{pmatrix}
6\\
3\\
2
\end{pmatrix}
\,, \qquad n_i\in \mathbbm{N}_0 \quad \forall i\,.
\end{equation}
We have discussed how to find solutions to such homogeneous Diophantine equation systems in section \ref{sec:chen}. After determining a Hilbert basis \(\mathcal{H}\) one can further split it to obtain physically viable solutions. Elements with \(n_1=1\) are assigned to \(\mathcal{H}_{\text{inhom}}\subset \mathcal{H}\) and basis elements with \(n_1=0\) are assigned to \(\mathcal{H}_{\text{hom}}\subset \mathcal{H}\). All physical solutions to equation (\ref{equa:hom}) are then given by the vectors
\begin{equation}
x=x_{\text{inhom}}(1+x_{\text{hom}}+x_{\text{hom}}^2+\ldots)\,,
\end{equation}
where \(\ldots\) denote higher powers in \(x_{\text{hom}}\). One has to take all possible combinations of elements \(x_{\text{inhom}}\in\mathcal{H}_{\text{inhom}}\) and \(x_{\text{hom}}\in\mathcal{H}_{\text{hom}}\). For practical purposes it seems reasonable to truncate the solution at some finite order in the fields \(\phi_i\).

\subsection{The superpotential to all orders}

We can now determine all monomials which give \(W\) for the string model under consideration. We follow the approach of \cite{Kappl:2011vi} and view all string selection rules as gauge and discrete $\mathbbm{Z}_N$ symmetries. Note that the space group is an infinite discrete non-Abelian group. However for $\mathbbm{Z}_6$-II models, it is possible to rephrase it in terms of the finite discrete group\footnote{There is an additional rule for some exceptional cases in the \(\text{G}_2\) orbifold plane \cite{Buchmuller:2006ik}. We have not been able to write it as a discrete symmetry and checked the invariance of the basis elements separately case by case a posterior.}  \(\mathbbm{Z}_3\times\mathbbm{Z}_2\times\mathbbm{Z}_2^\prime\) \cite{Buchmuller:2006ik}. 

Our model enjoys the symmetries \(\mathbbm{Z}^R_6\times \mathbbm{Z}^R_3\times \mathbbm{Z}^R_2\times \mathbbm{Z}_6\times\mathbbm{Z}_3\times\mathbbm{Z}_2\times \mathbbm{Z}_2^\prime\times\text{U(1)}^8\). The $R$ symmetries are remnants of the internal Lorentz symmetries, the \(\mathbbm{Z}_6\) symmetry results from the so-called point group selection rule, whereas the space group selection rule results in \(\mathbbm{Z}_3\times\mathbbm{Z}_2\times\mathbbm{Z}_2^\prime\). In addition we have to take care of the gauge symmetries resulting from the \(\text{E}_8\times\text{E}_8\). The non-Abelian symmetries can be discussed along the lines of \cite{Kappl:2011vi} and we focus on the Cartan generators leading to the \(\text{U}(1)\) factors given here. With the discussion of the $R$ symmetries from section \ref{sec:rsym} we get the Diophantine equation system
\begin{equation}
\begin{pmatrix}
\tilde{R}^1& R_4^1&\ldots&R_{M+3}^1\\
\tilde{R}^2& R_4^2&\ldots&R_{M+3}^2\\
\tilde{R}^3& R_4^3&\ldots&R_{M+3}^3\\
q_1^{\mathbbm{Z}_6}+q_2^{\mathbbm{Z}_6}+q_3^{\mathbbm{Z}_6}&q_4^{\mathbbm{Z}_6}&\ldots&q_{M+3}^{\mathbbm{Z}_6}\\
q_1^{\mathbbm{Z}_3}+q_2^{\mathbbm{Z}_3}+q_3^{\mathbbm{Z}_3}&q_4^{\mathbbm{Z}_3}&\ldots&q_{M+3}^{\mathbbm{Z}_3}\\
q_1^{\mathbbm{Z}_2}+q_2^{\mathbbm{Z}_2}+q_3^{\mathbbm{Z}_2}&q_4^{\mathbbm{Z}_2}&\ldots&q_{M+3}^{\mathbbm{Z}_2}\\
q_1^{\mathbbm{Z}_2^\prime}+q_2^{\mathbbm{Z}_2^\prime}+q_3^{\mathbbm{Z}_2^\prime}&q_4^{\mathbbm{Z}_2^\prime}&\ldots&q_{M+3}^{\mathbbm{Z}_2^\prime}\\
q_1^1+q_2^1+q_3^1&q_4^1&\ldots&q_{M+3}^1\\
\vdots&&&\vdots\\
q_1^8+q_2^8+q_3^8&q_4^8&\ldots&q_{M+3}^8\\
\end{pmatrix}
\cdot
\begin{pmatrix}
n_1\\
n_4\\
\vdots\\
n_{M+3}
\end{pmatrix}
=
\begin{pmatrix}
0\\
0\\
0\\
0\\
0\\
0\\
0\\
0\\
\vdots\\
0
\end{pmatrix}
\mod
\begin{pmatrix}
6\\
3\\
2\\
6\\
3\\
2\\
2\\
0\\
\vdots\\
0
\end{pmatrix}\,.
\end{equation}
We use \texttt{normaliz} \cite{normaliz:2014, 2012arXiv1206.1916B} to compute the Hilbert basis \(\mathcal{H}\) of this Diophantine equation system. Compared to other systems like \texttt{4ti2} \cite{4ti2} it has the advantage that the algorithm used is suited better for our problem \cite{Bruns20101098} and is also able to deal with congruences. We use the so-called "primal" algorithm \cite{Bruns20101098} which is much faster in our case then the so-called "dual" algorithm which is implemented in \texttt{4ti2}.

To make our results comparable with the results from the literature we reconsider benchmark model 1 from \cite{Lebedev:2007hv} in the following. We apply the rederived \(R\) symmetries which have been determined in \cite{Nilles:2013lda,Bizet:2013wha} and which we review in section \ref{sec:r}.
We consider a vacuum configuration in which we give the 14 standard model singlets
\begin{equation}
\label{equa:vev}
\tilde{s}=\{h_1, h_2, h_3, h_4, s_3^0, s_4^0, s_9^0, s_{10}^0, s_{12}^0, s_{21}^0, s_{24}^0, s_{28}^0, s_{29}^0, s_{30}^0\}
\end{equation}
a vacuum expectation value (VEV). To get the superpotential \(W\) of these fields to all orders, we can consider which fields induce the \(\mu\)-term. This is possible because \(\phi_1\bar{\phi}_1\) forms a complete singlet under all symmetries in this model \cite{Lebedev:2007hv,Kappl:2008ie,Brummer:2010fr}.
We find that up to order 10 in standard model singlets \(\tilde{s}_i\) the superpotential is given by
\begin{equation}
\begin{split}
\label{equa:super}
W&=(M_{1,\text{inhom}}+M_{2,\text{inhom}}+M_{3,\text{inhom}})(1+M_{1,\text{hom}}+M_{2,\text{hom}})\\
&=((s_3^0s_4^0+s_9^0s_{10}^0)s_{29}^0+h_3h_4s_{21}^0s_{31}^0)(1+
h_1h_2s_{21}^0s_{24}^0s_{28}^0s_{31}^0+(s_{21}^0s_{31}^0)^3)\,,
\end{split}
\end{equation}
with
\begin{equation}
\begin{aligned}
M_{1,\text{inhom}}&=s_3^0s_4^0s_{29}^0\,,&
M_{1,\text{hom}}&=(s_{21}^0s_{31}^0)^3\,,\\
M_{2,\text{inhom}}&=s_9^0s_{10}^0s_{29}^0,&
M_{2,\text{hom}}&=h_1h_2s_{21}^0s_{24}^0s_{28}^0s_{31}^0\,,\\
M_{3,\text{inhom}}&=h_3h_4s_{21}^0s_{31}^0\,.
\end{aligned}
\end{equation}
The monomials \(M_{i,\text{inhom}}\) are given by their exponent vectors \(x_{i,\text{inhom}}\) which are elements of the corresponding part of the Hilbert basis \(x_{i,\text{inhom}}\in\mathcal{H}_{\text{inhom}}\). The same is true for the homogeneous part. In principle we know all Hilbert basis elements and therefore all monomials which means we know the exact form of \(W\) to all orders.
We also see the manifestation of the \(\text{D}_4\) symmetry \cite{Ko:2007dz,Kobayashi:2006wq} present in this model which relates \(M_{1,\text{inhom}}\) and \(M_{2,\text{inhom}}\). There are two doublets
\begin{equation}
\tilde{D}_1=(s_3^0,s_9^0)\,,\qquad \tilde{D}_2=(s_4^0,s_{10}^0)\,,
\end{equation} 
which result in two invariants
\begin{equation}
\tilde{D}_1\cdot\tilde{D}_2^T s_{29}^0=(s_3^0s_4^0+s_9^0s_{10}^0)s_{29}^0=
M_{1,\text{inhom}}+M_{2,\text{inhom}}\,.
\end{equation}

\section{$\boldsymbol{R}$ symmetries in heterotic orbifolds}
\label{sec:r}

In this section we briefly review the origin of discrete $R$ symmetries in heterotic orbifolds \cite{Nilles:2013lda,Bizet:2013gf,Bizet:2013wha}. These symmetries arise from automorphisms of the orbifold space group. For the case of the $\mathbbm{Z}_6$-II orbifold the generators $R^\alpha$ can be written in terms of the Cartan generators of the internal $\mathrm{SO}(6)$ by means of twist vectors $v_\alpha$ according to $R^\alpha=\mathrm{e}^{2\pi \mathrm{i}(v^1_\alpha J_{45} + v^2_\alpha J_{67}+v^3_\alpha J_{78})}$. They are given by
\begin{equation}
\label{eq:RSymGenerators}
v_1 = \left(\tfrac{1}{6},\,0,\,0\right)\,,\hspace{3mm}v_2 = \left(0,\,\tfrac{1}{3},\,0\right)\,,\hspace{3mm}v_3 = \left(0,\,0,\,\tfrac{1}{2}\right)\,.
\end{equation}
In order to determine the charges of the corresponding fields with respect to the $R$ symmetries, the transformation behavior of the string states under these generators needs to be worked out.

String states of heterotic orbifolds are characterized by their left- and right-moving momenta $p_\mathrm{sh}$ and $q_\mathrm{sh}$, their left-moving oscillator excitations $\tilde{\alpha}$ and their locus. As the properties of the strings, described by these quantum numbers, are mainly independent of each other it is customary to write such states as
\begin{equation}
\label{eq:StringState}
\left|\Psi\right> = \tilde{\alpha}^{\mathcal{N}_\mathrm{L}}\,(\tilde{\alpha}^*)^{\bar{\mathcal{N}}_\mathrm{L}}\,\left|p_\mathrm{sh}\right> \otimes \left|q_\mathrm{sh}\right>\otimes \left|\text{locus}\right>\,.
\end{equation}
Here $\mathcal{N}_\mathrm{L}$ and $\bar{\mathcal{N}}_\mathrm{L}$ count the number of holomorphic and anti-holomorphic oscillators of the state. Recall that twisted strings have their center of mass momentum localized at fixed points of the space group $S$. The constructing element $g\in S$ of a fixed point $z_f$ is defined by $g z_f = z_f$. Untwisted strings can propagate freely and hence the factor describing the localization in \eqref{eq:StringState} is absent for such states. 

Note that we know the action of the generators \eqref{eq:RSymGenerators} on the covering space $\mathbbm{C}^3$ of the orbifold only. However, on $\mathbbm{C}^3$, there are infinitely many copies of each fixed point, which are related to $z_f$ by the action of space group elements $h$ that do not commute with $g$. The corresponding constructing elements form the conjugacy class $[g]=\{hgh^{-1}|h\in S\}$ of $g$. Therefore, if we denote the locus of a string state located at the fixed point $z_f$ with constructing element $g$ by $\left|g\right>$, a physical state corresponds to the linear combination
\begin{equation}
\label{eq:LocusLC}
\left|\text{locus}\right> = \sum_{g^\prime\in [g]} \mathrm{e}^{2\pi \mathrm{i} \tilde{\gamma}(g^\prime)}\left|\Psi_{g^\prime}\right>\,.
\end{equation}
The phases $\tilde{\gamma}$ are fixed by the requirement of invariance of the string states under elements $h$ that do not commute with $g$, such that 
\begin{equation}
\gamma(g,h) = \tilde{\gamma}(g) - \tilde{\gamma}(hgh^{-1}) = -p_\mathrm{sh}\cdot V_h + v_h\cdot (q_\mathrm{sh}-\mathcal{N}_\mathrm{L}+\bar{\mathcal{N}}_\mathrm{L}) \mod 1\,.
\end{equation}
Here $v_h$ denotes the twist vector of the space group element $h$ and $V_h$ denotes its gauge embedding.

The crucial observation is now, that the rotations \eqref{eq:RSymGenerators} map each space group element $g$ to a conjugate one, i.e.\ for each $g$ there exists an $h_g$ such that $R(g) = h_g\, g\, h_g^{-1}$. Therefore,the transformation behavior of the linear combination \eqref{eq:LocusLC} is given by 
\begin{equation}
\left|\text{locus}\right> \stackrel{R}{\longmapsto} \mathrm{e}^{-2\pi\mathrm{i} \gamma(g,h_g)}\left|\text{locus}\right>\,.
\end{equation}
For the other parts of the string states the transformation behavior follows from the transformation of the space time coordinates. A string state \eqref{eq:StringState} transforms according to
\begin{equation}
\left|\Psi\right> \stackrel{R}{\longmapsto} \exp\left[2\pi\mathrm{i}\, v\cdot\left(q_\mathrm{sh}-\mathcal{N}_\mathrm{L}+\bar{\mathcal{N}}_\mathrm{L} \right) -2\pi\mathrm{i}\, \gamma(g,h_g) \right]\left|\Psi\right>
\end{equation}
under an $R$ symmetry generator $R$ with shift vector $v$. For the case we are considering here, asking string correlators corresponding to superpotential terms to transform trivially under the $R$ symmetries one arrives at the $R$ charge selection rules \cite{Nilles:2013lda,Bizet:2013wha}
\begin{equation}
\label{eq:RSymmetriesZ6}
\begin{aligned}
\sum_{i=1}^L R^1_i = \sum_{i=1}^L\left[ q_{\mathrm{sh}\,i}^{\text{(boson)}\,1} - \mathcal{N}_{\mathrm{L}\,i}^1+\bar{\mathcal{N}}_{\mathrm{L}\,i}^1 - 6\gamma(g_i,\,h_{g_i})\right] &= -1 \mod 6\,,\\
\sum_{i=1}^L R^2_i =\sum_{i=1}^L\left[ q_{\mathrm{sh}\,i}^{\text{(boson)}\,2} - \mathcal{N}_{\mathrm{L}\,i}^2+\bar{\mathcal{N}}_{\mathrm{L}\,i}^2 -   3\gamma(g_i,\,h_{g_i})\right] &= -1 \mod 3\,,\\
\sum_{i=1}^L R^3_i =\sum_{i=1}^L\left[ q_{\mathrm{sh}\,i}^{\text{(boson)}\,3} - \mathcal{N}_{\mathrm{L}\,i}^3-\bar{\mathcal{N}}_{\mathrm{L}\,i}^3 - 2\gamma(g_i,\,h_{g_i})\right] &= -1 \mod 2\,.
\end{aligned}
\end{equation}
Here we have made use of the fact that the internal right-moving momenta $q_\mathrm{sh}$ of bosons and fermions within the same chiral multiplet are related by a shift of $\tfrac{1}{2}$. We denote the momentum of the respective bosons by $q_{\mathrm{sh}}^{\text{(boson)}}$ and define the $R$ charge of a string state to be that of the bosonic component of the chiral multiplet.

We modified the program \texttt{orbifolder} \cite{Nilles:2011aj} to incorporate the outlined \(R\) symmetries. The result for a concrete model can be found in appendix \ref{app:Spectrum}. 

\subsection{Differences to previous results}

Differences due to the new $R$ symmetries arise already at order 3 in the superpotential. Several new terms appear which were forbidden by the old rules. In the following we show a simple example. We look at the allowed couplings in the so-called benchmark model 1A of \cite{Lebedev:2007hv}. The $R$ charges of the considered fields are displayed in table \ref{tab:rcharges}. We immediately see that the coupling \(s_{13}^0s_{14}^0s_{30}^0\) was forbidden by the old rules but is allowed under the new $R$ symmetries.
\begin{table}[hbt]
\centering
\begin{tabular}{ccccccc}
&$R^{1,\text{old}}$&$R^{2,\text{old}}$&$R^{3,\text{old}}$&$R^{1,\text{new}}$&
$R^{2,\text{new}}$&$R^{3,\text{new}}$\\
\hline
$s_{13}^0$&$\frac{11}{6}$&$-\frac{1}{3}$&$-\frac{1}{2}$&$\frac{11}{6}$&$-\frac{1}{3}$&$-\frac{1}{2}$\Tspace\\
$s_{14}^0$&$\frac{11}{6}$&$-\frac{1}{3}$&$-\frac{1}{2}$&$\frac{11}{6}$&$-\frac{1}{3}$&$-\frac{1}{2}$\Tspace\\
$s_{30}^0$&$-\frac{5}{3}$&$-\frac{1}{3}$&$0$&$\frac{4}{3}$&$-\frac{1}{3}$&$0$\Tspace\\
\end{tabular}
\caption{Differences between the $R$ charges for some singlet fields.}
\label{tab:rcharges}
\end{table}

This has important phenomenological implications. If all fields \(s_{13}^0\), \(s_{14}^0\) and \(s_{30}^0\) get a VEV in a chosen configuration, the superpotential \(W\) also gets a VEV. As already discussed, the superpotential in such models is linked to the 
\(\mu\)-term. Therefore, in this configuration, the \(\mu\)-term will be generically too large. We will discuss this issue for our VEV configuration later in more detail.
There are also couplings forbidden by the new $R$ symmetries which were allowed by the old symmetries. The first example occurs at order 4 in the superpotential. 

\section{Phenomenological properties}
\label{sec:pheno}

In this section we want to briefly comment on the phenomenological properties of the considered model. Instead of doing a complete scan over different vacua we stick to benchmark model 1  (the field content of this model can be found in appendix \ref{app:Spectrum}). A full scan over different configurations is beyond the scope of this work.
Using the vacuum configuration given in equation (\ref{equa:vev}) we obtain the following features.

\subsection{$\boldsymbol{F}$-flatness, Gr\"obner bases and $\boldsymbol{D}$-flatness}

To analyze \(F\)-flatness we use techniques and methods known in computational algebraic geometry and discussed in high energy physics for example in \cite{Maniatis:2006jd} and in string theory in \cite{Gray:2006gn,Gray:2007yq}. We are looking for solutions to the equation system 
\begin{equation}
F_i=\frac{\partial W}{\partial \tilde{s}_i}=0, \qquad \forall i\,.
\end{equation}
To find a solution we truncate the superpotential \(W\) at a given order. We take \(W\) up to order 10 in standard model singlets which was given in equation (\ref{equa:super}) and set all coupling coefficients to unity for simplicity. In principle these coefficients are calculable functions of the geometric moduli of the compactification. This dependence can be used to stabilize the geometric moduli \cite{Brummer:2010fr}.

We use \texttt{Singular} \cite{dgps} to compute the Gr\"obner basis of the ideal generated by the \(F\)-term equations. Afterwards we compute the primary decomposition and search for \(F\)-flat solutions \(F_i=0\). We find only one non-trivial branch of solutions. Trivial solutions \(\left<\tilde{s}_i\right>=0\), would violate our assumption that all fields \(\tilde{s}_i\) given in equation (\ref{equa:vev}) obtain a non-vanishing VEV. The only non-trivial branch can be solved, for example, by 
\begin{equation}
\left<s_{28}^0\right>=-\frac{1+\left<s_{21}^0\right>^3\left<s_{31}^0\right>^3}
{\left<h_1\right>\left<h_2\right>\left<s_{21}^0\right>\left<s_{24}^0\right>\left<s_{31}^0\right>}\,,
\qquad
\left<s_{29}^0\right>=-\frac{\left<h_3\right>\left<h_4\right>\left<s_{21}^0\right>
\left<s_{31}^0\right>}{\left<s_3^0\right>\left<s_4^0\right>+\left<s_9^0\right>\left<s_{10}^0\right>}\,,
\end{equation}
which results in \(F_i=\left<W\right>=0\). That means we can solve all 14 \(F\)-term equations simultaneously by fixing only two VEVs. This is nearly the opposite behavior to the one discussed in \cite{Kappl:2010yu} where a remnant \(\mathbbm{Z}_4^R\) symmetry \cite{Lee:2010gv} has been used to restrict the superpotential\footnote{Further applications of this symmetry to flavor model building are discussed in \cite{Dreiner:2013ala}, whereas the relation to \(R\) parity violation is studied in \cite{Chen:2014gua}.}. There it seems to be more fertile to look for minima in which all singlets get fixed by the \(F\)-term equations. It is interesting to study how the symmetries of the superpotential determine the solution structure of the \(F\)-term equations. 
In our case many \(F\)-term equations are degenerate because of the remnant \(\text{D}_4\) symmetry. The necessary breaking of this symmetry at a lower scale (see section \ref{sec:yuka}) can therefore be potentially used to stabilize additional moduli. The detailed study of this mechanism is beyond the scope of this work. At this stage we are satisfied with finding a consistent, non-trivial solution.

We also would like to emphasize that the \(\mu\)-term in the minima determined above, vanishes because \(\left<W\right>=0\). This is well known to be related to approximate $R$ symmetries \cite{Kappl:2008ie}. As we know the superpotential not only to order 10 in singlets, but to all orders, it seems to be possible to address the question of \(F\)-flatness to all orders. It is effective to attack this problem in terms of the Hilbert basis monomials.
In this way, the given non-trivial solution can be extended to all orders. We can split the superpotential according to the Hilbert basis monomials into two pieces
\begin{equation}
W(\tilde{s}_i)=W_{\text{inhom}}(\tilde{s}_i)(1+W_{\text{hom}}(\tilde{s}_i))\,.
\end{equation}
Here \(W_{\text{inhom}}\) is given by the linear combination of all inhomogeneous Hilbert basis monomials \(M_{i,\text{inhom}}\). Furthermore \(W_{\text{hom}}\) denotes all possible combinations of homogeneous monomials \(M_{i,\text{hom}}\).
The \(F\)-term equations in this picture are given by
\begin{equation}
F_i=\frac{\partial W_{\text{inhom}}}{\partial \tilde{s}_i}(1+W_{\text{hom}})
+W_{\text{inhom}}\frac{\partial W_{\text{hom}}}{\partial \tilde{s}_i}\,, \qquad \forall i\,.
\end{equation}
It is obvious that all \(F\)-term equations vanish if \(W_{\text{inhom}}(\tilde{s}_i)=0\) and \(1+W_{\text{hom}}(\tilde{s}_i)=0\) simultaneously. Our concrete solution at order 10 is exactly of this kind. This behavior is not limited to any order and in general such a solution will exist for the superpotential at higher order. The disadvantage that generically only two fields are fixed in this solution branch nevertheless remains.
Different strategies to find minima like the ones considered in \cite{Camargo-Molina:2013qva,Cicoli:2013cha} may also help to address the computational difficulties.

Let us also comment on \(D\)-flatness. It is possible to satisfy \(D\)-flatness along the lines of \cite{Buccella:1982nx, Cleaver:1997jb}. We have explicitly checked that we can cancel the Fayet-Iliopoulos term by a monomial which carries negative charge under the anomalous \(\text{U}(1)\). \(D\)-flatness will also help to fix some VEVs (see for example \cite{Kappl:2010yu}) but in our concrete example some flat directions remain.

\subsection{Yukawa couplings}
\label{sec:yuka}

For the Yukawa interactions we obtain
\begin{equation}
W_{\text{Yuk}}=Y_u(\tilde{s}_i)q\bar{u}\bar{\phi}_1+Y_d(\tilde{s}_i)q\bar{d}\phi_1+
Y_e(\tilde{s}_i)l\bar{e}\phi_1\,,
\end{equation}
where the Yukawa matrices \(Y_i(\tilde{s}_i)\) depend on the singlet fields \(\tilde{s}_i\) to which we have assigned VEVs (see equation (\ref{equa:vev})). We computed for each coupling the corresponding Hilbert basis and found that, at lowest order in singlets, the structure is
\begin{equation}
Y_u=\begin{pmatrix}
M_1&M_2&M_3+M_4\\
M_2&M_1&M_5+M_6\\
M_7+M_8+M_9+M_{10}&M_{11}+M_{12}+M_{13}+M_{14}&1
\end{pmatrix}\,,
\end{equation}
with
\begin{equation}
\begin{aligned}
M_1&=h_1h_3s_4^0s_{21}^0s_{29}^0s_{31}^0\,,&
M_2&=h_1h_3s_{10}^0s_{21}^0s_{29}^0s_{31}^0\,,\\
M_3&=s_3^0M_2\,,&
M_4&=s_9^0M_1\,,\\
M_5&=s_3^0M_1\,,&
M_6&=s_9^0M_2\,,\\
M_7&=s_{12}^0M_1\,,&
M_8&=s_3^0s_4^0s_{10}^0s_{21}^0(s_{29}^0)^2s_{31}^0\,,\\
M_9&=s_9^0(s_{10}^0)^2s_{21}^0(s_{29}^0)^2s_{31}^0\,,&
M_{10}&=(s_4^0)^2s_9^0s_{21}^0(s_{29}^0)^2s_{31}^0\,,\\
M_{11}&=s_{12}^0M_2\,,&
M_{12}&=s_4^0s_9^0s_{10}^0s_{21}^0(s_{29}^0)^2s_{31}^0\,,\\
M_{13}&=s_3^0(s_{10}^0)^2s_{21}^0(s_{29}^0)^2s_{31}^0\,,&
M_{14}&=s_3^0(s_4^0)^2s_{21}^0(s_{29}^0)^2s_{31}^0\,.
\end{aligned}
\end{equation}
Further, for the down quarks and leptons
\begin{equation}
Y_d=\begin{pmatrix}
M_1&M_2&M_3+M_4\\
M_2&M_1&M_5\\
M_6&M_7&M_8+M_9
\end{pmatrix}
\,,\qquad
Y_e=\begin{pmatrix}
M_1&M_2&M_{10}+M_{11}\\
M_2&M_1&M_{12}\\
M_{13}&M_{14}&M_{15}+M_{16}
\end{pmatrix}\,,
\end{equation}
with
\begin{equation}
\begin{aligned}
M_1&=h_1h_2s_9^0s_{12}^0s_{29}^0\,,&
M_2&=h_1h_2s_3^0s_{12}^0s_{29}^0\,,\\
M_3&=s_9^0M_1\,,&
M_4&=s_3^0M_2\,,\\
M_5&=s_3^0M_1\,,&
M_6&=s_{12}^0s_{21}^0M_1\,,\\
M_7&=s_{12}^0s_{21}^0M_2\,,&
M_8&=s_9^0s_{12}^0s_{21}^0M_1\,,\\
M_9&=s_3^0s_{12}^0s_{21}^0M_2\,,&
M_{10}&=(s_9^0)^2s_{12}^0s_{21}^0s_{29}^0\,,\\
M_{11}&=(s_3^0)^2s_{12}^0s_{21}^0s_{29}^0\,,&
M_{12}&=s_3^0s_4^0s_{12}^0s_{21}^0s_{29}^0\,,\\
M_{13}&=s_{12}^0M_1\,,&
M_{14}&=s_{12}^0M_2\,,\\
M_{15}&=s_{12}^0M_{10}\,,&
M_{16}&=s_{12}^0M_{11}\,.
\end{aligned}
\end{equation}
Let us note that, as expected, the top quark coupling is of order one. The reason for this behavior is the connection of the coupling to the higher dimensional gauge coupling \cite{Hosteins:2009xk}. Thus a realistic top quark mass in this model is guaranteed.

As a consequence of the localization of the first two generations in the extra dimensional space, these fields form a doublet under the \(\text{D}_4\) symmetry \cite{Ko:2007dz,Kobayashi:2006wq}. This manifests itself in the appearance of the monomials $M_1$ and $M_2$ in the Yukawa matrices. However, this symmetry needs to be broken at a lower scale to explain the different masses between the first and second generation \cite{Kappl:2010sg}.

\subsection{Additional features}

Also with the new $R$ symmetries all exotics can be made massive. We will discuss one example in detail and skip the details for the other exotics. As can be seen from table \ref{tab:yfields} the \(R^3\) charges of \(y_1\) and \(y_2\) have changed under the new $R$ symmetry. 
\begin{table}[hbt]
\centering
\begin{tabular}{ccccccc}
&$R^{1,\text{old}}$&$R^{2,\text{old}}$&$R^{3,\text{old}}$&$R^{1,\text{new}}$&
$R^{2,\text{new}}$&$R^{3,\text{new}}$\\
\hline
$y_{1}$&$-\frac{1}{6}$&$-\frac{1}{3}$&$-\frac{1}{2}$&$-\frac{1}{6}$&$-\frac{1}{3}$&$\frac{1}{2}$\Tspace\\
$y_{2}$&$-\frac{1}{6}$&$-\frac{1}{3}$&$-\frac{1}{2}$&$-\frac{1}{6}$&$-\frac{1}{3}$&$\frac{1}{2}$\Tspace\\
\end{tabular}
\caption{Differences between the $R$ charges of \(y_1\) and \(y_2\).}
\label{tab:yfields}
\end{table}
Nevertheless the mass matrix does not change, because both fields always appear in pairs and we have
\begin{equation}
\sum_i R^3_i=-1\mod 2
\end{equation} 
and \(1 = -1 \mod 2\). We have found that all exotics get a high mass and therefore decouple from the low energy particle spectrum. 

The situation of the generation of neutrino masses remains the same also after including the new $R$ symmetries. Thus, neutrino masses can be generated with the see-saw mechanism for many right-handed neutrinos \cite{Buchmuller:2007zd,Heeck:2012fw}. 

Our results for proton decay do not differ substantially from the ones described in \cite{Lebedev:2007hv}. We find that \(qqql\) operators as well as couplings with massive exotic triplets like \(q_1l_1\bar{\delta}_4\) and \(q_1q_1\delta_4\) are allowed. After integrating out the exotic triplets the trilinear operators might be a further source of proton decay. A full clarification of these
questions needs a detailed examination of the vacuum configurations of benchmark model 1 and is
beyond the scope of this paper.

\section{Conclusions}

We have shown that with the recently rederived \(R\) symmetries the phenomenological attractive models of the mini-landscape remain viable. All appealing features survive and can be understood also from the viewpoint of Hilbert basis monomials. We have shown how symmetries lead to Diophantine equations and how one can find all solutions. It was possible to use this approach for a model in the mini-landscape and its large number of \(R\) and non-\(R\) symmetries. With this method we have been able to determine the coupling structure to all orders in singlet fields. As an example we have shown that this approach can also be useful for flavor model building and not only for string model building. The symmetries of a given model directly constrain the superpotential and its Hilbert basis building blocks. As has been shown, this can be useful in understanding the \(F\)-flatness conditions to find supersymmetry preserving vacua. More work in this direction seems to be interesting. Especially to find point-like minima where all fields and thus all moduli are stabilized. On the other hand the extension of the Hilbert basis approach to non-Abelian discrete symmetries for flavor model building seems to be desirable. An application of our approach to heterotic models based on different orbifold geometries like \(\mathbbm{Z}_2\times \mathbbm{Z}_2\) is another potential extension.  

\subsection*{Acknowledgments}

We want to thank Dami\'an Mayorga Pe\~na for discussions. This work was supported by the SFB-Transregio TR33 "The Dark Universe" (Deutsche Forschungsgemeinschaft).

\appendix

\section{Details of model 1}\label{app:Spectrum}
Model 1 is based on the gauge embedding 
\newlength{\SOFWM}
\settowidth{\SOFWM}{$\tfrac{9}{2}$}
\newcommand{\esfrac}[2]{\hphantom{-}\makebox[\SOFWM]{$\tfrac{#1}{#2}$}}
\newcommand{\esint}[1]{\hphantom{-}\makebox[\SOFWM]{$#1$}}
\begin{equation*}
\begin{aligned}
V &= (\esfrac{1}{3},-\tfrac{1}{2},-\tfrac{1}{2},\esint{0},\esint{0},\esint{0},\esint{0},\esint{0})(\esfrac{1}{2},-\tfrac{1}{6},-\tfrac{1}{2},-\tfrac{1}{2},-\tfrac{1}{2},-\tfrac{1}{2},-\tfrac{1}{2},\esfrac{1}{2})\,,\\
W_3 = W_4 &= (-\tfrac{1}{2},-\tfrac{1}{2},\esfrac{1}{6},\esfrac{1}{6},\esfrac{1}{6},\esfrac{1}{6},\esfrac{1}{6},\esfrac{1}{6})(\esfrac{1}{3},\esint{0},\esint{0},\esfrac{2}{3},\esint{0},\esfrac{5}{3},-\makebox[\SOFWM]{$2$},\esint{0})\,,\\
W_5 &= (\esint{0},-\tfrac{1}{2},-\tfrac{1}{2},-\tfrac{1}{2},\esfrac{1}{2},\esint{0},\esint{0},\esint{0})(\esint{4},-\makebox[\SOFWM]{$3$},-\tfrac{7}{2},-\makebox[\SOFWM]{$4$},-\makebox[\SOFWM]{$3$},-\tfrac{7}{2},-\tfrac{9}{2},\esfrac{7}{2})\,.
\end{aligned}
\end{equation*}

Here we list the complete spectrum of massless string states including their $R$ charges. Those $R$ charges that differ from the ``old'' ones are marked by red color.

\begin{onehalfspacing} \footnotesize \begingroup \setlength\tabcolsep{3pt}\begin{longtable}{ccccccccccccccccccccc}
label&$k$&$n_3$&$n_2$&$n_2^\prime$&$q_\gamma$&$\mathrm{R}^1$&$\mathrm{R}^2$&$\mathrm{R}^3$&representation&$q_Y$&$q^1$&$q^2$&$q^3$&$q^4$&$q^5$&$q^6$&$q^7$&$q^8$&$q_{\text{anom}}$&$q_\mathrm{B-L}$\\\hline \endhead
$\bar{n}_{3}$&$0$&$0$&$0$&$0$&$0$&$-1$&$0$&$0$&$\left(\mathbf{1},\,\mathbf{1},\,\mathbf{1},\,\mathbf{1}\right)$&$0$&$-\tfrac{1}{2}$&$-\tfrac{1}{2}$&$\tfrac{1}{2}$&$\tfrac{5}{2}$&$0$&$0$&$0$&$0$&$-\tfrac{1}{3}$&$-1$\\
$\bar{e}_{3}$&$0$&$0$&$0$&$0$&$0$&$-1$&$0$&$0$&$\left(\mathbf{1},\,\mathbf{1},\,\mathbf{1},\,\mathbf{1}\right)$&$-1$&$-\tfrac{1}{2}$&$\tfrac{1}{2}$&$-\tfrac{1}{2}$&$\tfrac{1}{2}$&$0$&$0$&$0$&$0$&$\tfrac{2}{3}$&$-1$\\
$\bar{u}_{3}$&$0$&$0$&$0$&$0$&$0$&$-1$&$0$&$0$&$\left(\mathbf{\bar{3}},\,\mathbf{1},\,\mathbf{1},\,\mathbf{1}\right)$&$\tfrac{2}{3}$&$-\tfrac{1}{2}$&$\tfrac{1}{2}$&$-\tfrac{1}{2}$&$\tfrac{1}{2}$&$0$&$0$&$0$&$0$&$\tfrac{2}{3}$&$\tfrac{1}{3}$\\
$\bar{f}_{1}$&$0$&$0$&$0$&$0$&$0$&$-1$&$0$&$0$&$\left(\mathbf{1},\,\mathbf{1},\,\mathbf{\bar{4}},\,\mathbf{1}\right)$&$0$&$0$&$0$&$0$&$0$&$-\tfrac{1}{2}$&$1$&$\tfrac{1}{2}$&$\tfrac{1}{2}$&$\tfrac{5}{3}$&$1$\\
$f_{1}$&$0$&$0$&$0$&$0$&$0$&$-1$&$0$&$0$&$\left(\mathbf{1},\,\mathbf{1},\,\mathbf{4},\,\mathbf{1}\right)$&$0$&$0$&$0$&$0$&$0$&$-\tfrac{1}{2}$&$-1$&$\tfrac{1}{2}$&$-\tfrac{1}{2}$&$\tfrac{2}{3}$&$-1$\\
$\phi_{1}$&$0$&$0$&$0$&$0$&$0$&$0$&$0$&$-1$&$\left(\mathbf{1},\,\mathbf{2},\,\mathbf{1},\,\mathbf{1}\right)$&$\tfrac{1}{2}$&$0$&$0$&$-1$&$1$&$0$&$0$&$0$&$0$&$2$&$0$\\
$\bar{\phi}_{1}$&$0$&$0$&$0$&$0$&$0$&$0$&$0$&$-1$&$\left(\mathbf{1},\,\mathbf{2},\,\mathbf{1},\,\mathbf{1}\right)$&$-\tfrac{1}{2}$&$0$&$0$&$1$&$-1$&$0$&$0$&$0$&$0$&$-2$&$0$\\
$s^0_{2}$&$0$&$0$&$0$&$0$&$0$&$0$&$-1$&$0$&$\left(\mathbf{1},\,\mathbf{1},\,\mathbf{1},\,\mathbf{1}\right)$&$0$&$0$&$0$&$0$&$0$&$-1$&$0$&$-1$&$0$&$-\tfrac{5}{3}$&$0$\\
$s^0_{1}$&$0$&$0$&$0$&$0$&$0$&$0$&$-1$&$0$&$\left(\mathbf{1},\,\mathbf{1},\,\mathbf{1},\,\mathbf{1}\right)$&$0$&$0$&$0$&$0$&$0$&$-1$&$0$&$1$&$0$&$\tfrac{7}{3}$&$0$\\
$q_{3}$&$0$&$0$&$0$&$0$&$0$&$0$&$-1$&$0$&$\left(\mathbf{3},\,\mathbf{2},\,\mathbf{1},\,\mathbf{1}\right)$&$-\tfrac{1}{6}$&$\tfrac{1}{2}$&$-\tfrac{1}{2}$&$-\tfrac{1}{2}$&$\tfrac{1}{2}$&$0$&$0$&$0$&$0$&$\tfrac{4}{3}$&$-\tfrac{1}{3}$\\
\hline
$n_{12}$&$2$&$0$&$0$&$0$&$0$&$-\tfrac{2}{3}$&${\color{red}\tfrac{2}{3}\normalcolor}$&$0$&$\left(\mathbf{1},\,\mathbf{1},\,\mathbf{1},\,\mathbf{1}\right)$&$0$&$-\tfrac{5}{6}$&$\tfrac{1}{2}$&$-\tfrac{1}{6}$&$-\tfrac{5}{6}$&$-\tfrac{1}{3}$&$\tfrac{2}{3}$&$0$&$\tfrac{1}{3}$&$-\tfrac{1}{9}$&$1$\\
$\bar{f}_{4}$&$2$&$0$&$0$&$0$&$0$&$-\tfrac{2}{3}$&${\color{red}\tfrac{2}{3}\normalcolor}$&$0$&$\left(\mathbf{1},\,\mathbf{1},\,\mathbf{\bar{4}},\,\mathbf{1}\right)$&$0$&$\tfrac{1}{6}$&$-\tfrac{1}{2}$&$-\tfrac{1}{6}$&$-\tfrac{5}{6}$&$\tfrac{1}{6}$&$-\tfrac{1}{3}$&$\tfrac{1}{2}$&$-\tfrac{1}{6}$&$\tfrac{8}{9}$&$0$\\
$\delta_{6}$&$2$&$0$&$0$&$0$&$\tfrac{1}{2}$&${\color{red}\tfrac{7}{3}\normalcolor}$&$-\tfrac{1}{3}$&$0$&$\left(\mathbf{3},\,\mathbf{1},\,\mathbf{1},\,\mathbf{1}\right)$&$\tfrac{1}{3}$&$-\tfrac{1}{3}$&$0$&$\tfrac{1}{3}$&$\tfrac{2}{3}$&$-\tfrac{1}{3}$&$\tfrac{2}{3}$&$0$&$\tfrac{1}{3}$&$-\tfrac{1}{9}$&$\tfrac{2}{3}$\\
$\bar{n}_{9}$&$2$&$0$&$0$&$0$&$\tfrac{1}{2}$&${\color{red}\tfrac{7}{3}\normalcolor}$&$-\tfrac{1}{3}$&$0$&$\left(\mathbf{1},\,\mathbf{1},\,\mathbf{1},\,\mathbf{1}\right)$&$0$&$\tfrac{1}{6}$&$-\tfrac{1}{2}$&$-\tfrac{1}{6}$&$-\tfrac{5}{6}$&$\tfrac{2}{3}$&$\tfrac{2}{3}$&$0$&$-\tfrac{2}{3}$&$-\tfrac{1}{9}$&$-1$\\
$\bar{\eta}_{3}$&$2$&$0$&$0$&$0$&$\tfrac{1}{2}$&${\color{red}\tfrac{7}{3}\normalcolor}$&${\color{red}-\tfrac{4}{3}\normalcolor}$&$0$&$\left(\mathbf{1},\,\mathbf{1},\,\mathbf{1},\,\mathbf{2}\right)$&$0$&$\tfrac{1}{6}$&$-\tfrac{1}{2}$&$-\tfrac{1}{6}$&$-\tfrac{5}{6}$&$-\tfrac{1}{3}$&$-\tfrac{1}{3}$&$0$&$-\tfrac{2}{3}$&$-\tfrac{1}{9}$&$-1$\\
$\bar{d}_{3}$&$2$&$0$&$0$&$0$&$0$&$-\tfrac{2}{3}$&${\color{red}-\tfrac{4}{3}\normalcolor}$&$0$&$\left(\mathbf{\bar{3}},\,\mathbf{1},\,\mathbf{1},\,\mathbf{1}\right)$&$-\tfrac{1}{3}$&$\tfrac{1}{6}$&$\tfrac{1}{2}$&$-\tfrac{1}{6}$&$\tfrac{1}{6}$&$-\tfrac{1}{3}$&$\tfrac{2}{3}$&$0$&$\tfrac{1}{3}$&$\tfrac{8}{9}$&$\tfrac{1}{3}$\\
$s^0_{31}$&$2$&$0$&$0$&$0$&$\tfrac{1}{2}$&${\color{red}\tfrac{7}{3}\normalcolor}$&${\color{red}\tfrac{2}{3}\normalcolor}$&$0$&$\left(\mathbf{1},\,\mathbf{1},\,\mathbf{1},\,\mathbf{1}\right)$&$0$&$\tfrac{2}{3}$&$0$&$\tfrac{1}{3}$&$\tfrac{5}{3}$&$-\tfrac{1}{3}$&$\tfrac{2}{3}$&$0$&$\tfrac{1}{3}$&$\tfrac{8}{9}$&$0$\\
\hline
$\delta_{4}$&$2$&$0$&$0$&$0$&$0$&$-\tfrac{2}{3}$&$-\tfrac{1}{3}$&$0$&$\left(\mathbf{3},\,\mathbf{1},\,\mathbf{1},\,\mathbf{1}\right)$&$\tfrac{1}{3}$&$-\tfrac{1}{3}$&$0$&$0$&$-1$&$\tfrac{2}{3}$&$0$&$0$&$0$&$-\tfrac{7}{9}$&$\tfrac{2}{3}$\\
$h_{8}$&$2$&$0$&$0$&$0$&$0$&$-\tfrac{2}{3}$&$-\tfrac{1}{3}$&$0$&$\left(\mathbf{1},\,\mathbf{1},\,\mathbf{1},\,\mathbf{2}\right)$&$0$&$\tfrac{2}{3}$&$0$&$0$&$0$&$-\tfrac{1}{3}$&$-1$&$0$&$0$&$\tfrac{2}{9}$&$0$\\
$\bar{\delta}_{4}$&$2$&$0$&$0$&$0$&$0$&$-\tfrac{2}{3}$&$-\tfrac{1}{3}$&$0$&$\left(\mathbf{\bar{3}},\,\mathbf{1},\,\mathbf{1},\,\mathbf{1}\right)$&$-\tfrac{1}{3}$&$-\tfrac{1}{3}$&$0$&$0$&$1$&$\tfrac{2}{3}$&$0$&$0$&$0$&$-\tfrac{1}{9}$&$-\tfrac{2}{3}$\\
$h_{7}$&$2$&$0$&$0$&$0$&$0$&$-\tfrac{2}{3}$&$-\tfrac{1}{3}$&$0$&$\left(\mathbf{1},\,\mathbf{1},\,\mathbf{1},\,\mathbf{2}\right)$&$0$&$\tfrac{2}{3}$&$0$&$0$&$0$&$-\tfrac{1}{3}$&$1$&$0$&$0$&$\tfrac{8}{9}$&$0$\\
$s^0_{25}$&$2$&$0$&$0$&$0$&$0$&$-\tfrac{2}{3}$&$-\tfrac{1}{3}$&$0$&$\left(\mathbf{1},\,\mathbf{1},\,\mathbf{1},\,\mathbf{1}\right)$&$0$&$\tfrac{2}{3}$&$0$&$0$&$0$&$-\tfrac{1}{3}$&$0$&$-1$&$0$&$-\tfrac{13}{9}$&$0$\\
$s^0_{24}$&$2$&$0$&$0$&$0$&$0$&$-\tfrac{2}{3}$&$-\tfrac{1}{3}$&$0$&$\left(\mathbf{1},\,\mathbf{1},\,\mathbf{1},\,\mathbf{1}\right)$&$0$&$\tfrac{2}{3}$&$0$&$0$&$0$&$-\tfrac{1}{3}$&$0$&$1$&$0$&$\tfrac{23}{9}$&$0$\\
$s^0_{30}$&$2$&$0$&$0$&$0$&$\tfrac{1}{2}$&${\color{red}\tfrac{4}{3}\normalcolor}$&$-\tfrac{1}{3}$&$0$&$\left(\mathbf{1},\,\mathbf{1},\,\mathbf{1},\,\mathbf{1}\right)$&$0$&$\tfrac{2}{3}$&$0$&$0$&$0$&$\tfrac{2}{3}$&$0$&$0$&$0$&$\tfrac{2}{9}$&$0$\\
$s^0_{26}$&$2$&$0$&$0$&$0$&$0$&$-\tfrac{2}{3}$&$\tfrac{2}{3}$&$0$&$\left(\mathbf{1},\,\mathbf{1},\,\mathbf{1},\,\mathbf{1}\right)$&$0$&$\tfrac{2}{3}$&$0$&$0$&$0$&$\tfrac{2}{3}$&$0$&$0$&$0$&$\tfrac{2}{9}$&$0$\\
\hline
$\bar{\delta}_{6}$&$2$&$2$&$0$&$0$&$\tfrac{1}{2}$&${\color{red}\tfrac{7}{3}\normalcolor}$&${\color{red}\tfrac{2}{3}\normalcolor}$&$0$&$\left(\mathbf{\bar{3}},\,\mathbf{1},\,\mathbf{1},\,\mathbf{1}\right)$&$-\tfrac{1}{3}$&$-\tfrac{1}{3}$&$0$&$-\tfrac{1}{3}$&$-\tfrac{2}{3}$&$-\tfrac{1}{3}$&$-\tfrac{2}{3}$&$0$&$-\tfrac{1}{3}$&$-\tfrac{1}{9}$&$-\tfrac{2}{3}$\\
$\bar{n}_{11}$&$2$&$2$&$0$&$0$&$0$&$-\tfrac{2}{3}$&$-\tfrac{1}{3}$&$0$&$\left(\mathbf{1},\,\mathbf{1},\,\mathbf{1},\,\mathbf{1}\right)$&$0$&$\tfrac{1}{6}$&$-\tfrac{1}{2}$&$\tfrac{1}{6}$&$\tfrac{5}{6}$&$-\tfrac{1}{3}$&$\tfrac{4}{3}$&$0$&$-\tfrac{1}{3}$&$\tfrac{5}{9}$&$-1$\\
$\bar{n}_{10}$&$2$&$2$&$0$&$0$&$0$&$-\tfrac{2}{3}$&$-\tfrac{1}{3}$&$0$&$\left(\mathbf{1},\,\mathbf{1},\,\mathbf{1},\,\mathbf{1}\right)$&$0$&$\tfrac{1}{6}$&$\tfrac{1}{2}$&$-\tfrac{5}{6}$&$\tfrac{5}{6}$&$-\tfrac{1}{3}$&$-\tfrac{2}{3}$&$0$&$-\tfrac{1}{3}$&$\tfrac{14}{9}$&$-1$\\
$\bar{\eta}_{4}$&$2$&$2$&$0$&$0$&$0$&$-\tfrac{2}{3}$&${\color{red}\tfrac{2}{3}\normalcolor}$&$0$&$\left(\mathbf{1},\,\mathbf{1},\,\mathbf{1},\,\mathbf{2}\right)$&$0$&$\tfrac{1}{6}$&$-\tfrac{1}{2}$&$\tfrac{1}{6}$&$\tfrac{5}{6}$&$\tfrac{2}{3}$&$\tfrac{1}{3}$&$0$&$-\tfrac{1}{3}$&$-\tfrac{1}{9}$&$-1$\\
$\bar{f}_{6}$&$2$&$2$&$0$&$0$&$\tfrac{1}{2}$&${\color{red}\tfrac{7}{3}\normalcolor}$&${\color{red}-\tfrac{4}{3}\normalcolor}$&$0$&$\left(\mathbf{1},\,\mathbf{1},\,\mathbf{\bar{4}},\,\mathbf{1}\right)$&$0$&$\tfrac{1}{6}$&$-\tfrac{1}{2}$&$\tfrac{1}{6}$&$\tfrac{5}{6}$&$\tfrac{1}{6}$&$\tfrac{1}{3}$&$-\tfrac{1}{2}$&$\tfrac{1}{6}$&$-\tfrac{7}{9}$&$0$\\
$s^0_{32}$&$2$&$2$&$0$&$0$&$\tfrac{1}{2}$&${\color{red}\tfrac{7}{3}\normalcolor}$&${\color{red}-\tfrac{4}{3}\normalcolor}$&$0$&$\left(\mathbf{1},\,\mathbf{1},\,\mathbf{1},\,\mathbf{1}\right)$&$0$&$\tfrac{2}{3}$&$0$&$-\tfrac{1}{3}$&$-\tfrac{5}{3}$&$-\tfrac{1}{3}$&$-\tfrac{2}{3}$&$0$&$-\tfrac{1}{3}$&$\tfrac{2}{9}$&$0$\\
$\bar{l}_{1}$&$2$&$2$&$0$&$0$&$\tfrac{1}{2}$&${\color{red}\tfrac{7}{3}\normalcolor}$&$-\tfrac{1}{3}$&$0$&$\left(\mathbf{1},\,\mathbf{2},\,\mathbf{1},\,\mathbf{1}\right)$&$-\tfrac{1}{2}$&$\tfrac{1}{6}$&$\tfrac{1}{2}$&$\tfrac{1}{6}$&$-\tfrac{1}{6}$&$-\tfrac{1}{3}$&$-\tfrac{2}{3}$&$0$&$-\tfrac{1}{3}$&$-\tfrac{4}{9}$&$-1$\\
$\bar{n}_{16}$&$2$&$2$&$0$&$0$&$\tfrac{1}{2}$&${\color{red}\tfrac{4}{3}\normalcolor}$&${\color{red}-\tfrac{4}{3}\normalcolor}$&$0$&$\left(\mathbf{1},\,\mathbf{1},\,\mathbf{1},\,\mathbf{1}\right)$&$0$&$\tfrac{1}{6}$&$-\tfrac{1}{2}$&$\tfrac{1}{6}$&$\tfrac{5}{6}$&$-\tfrac{1}{3}$&$-\tfrac{2}{3}$&$0$&$-\tfrac{1}{3}$&$-\tfrac{1}{9}$&$-1$\\
$\bar{n}_{12}$&$2$&$2$&$0$&$0$&$0$&$-\tfrac{2}{3}$&${\color{red}-\tfrac{1}{3}\normalcolor}$&$0$&$\left(\mathbf{1},\,\mathbf{1},\,\mathbf{1},\,\mathbf{1}\right)$&$0$&$\tfrac{1}{6}$&$-\tfrac{1}{2}$&$\tfrac{1}{6}$&$\tfrac{5}{6}$&$-\tfrac{1}{3}$&$-\tfrac{2}{3}$&$0$&$-\tfrac{1}{3}$&$-\tfrac{1}{9}$&$-1$\\
\hline
$n_{13}$&$2$&$0$&$0$&$0$&$1$&$-\tfrac{2}{3}$&${\color{red}\tfrac{2}{3}\normalcolor}$&$0$&$\left(\mathbf{1},\,\mathbf{1},\,\mathbf{1},\,\mathbf{1}\right)$&$0$&$-\tfrac{5}{6}$&$\tfrac{1}{2}$&$-\tfrac{1}{6}$&$-\tfrac{5}{6}$&$-\tfrac{1}{3}$&$\tfrac{2}{3}$&$0$&$\tfrac{1}{3}$&$-\tfrac{1}{9}$&$1$\\
$\bar{f}_{5}$&$2$&$0$&$0$&$0$&$1$&$-\tfrac{2}{3}$&${\color{red}\tfrac{2}{3}\normalcolor}$&$0$&$\left(\mathbf{1},\,\mathbf{1},\,\mathbf{\bar{4}},\,\mathbf{1}\right)$&$0$&$\tfrac{1}{6}$&$-\tfrac{1}{2}$&$-\tfrac{1}{6}$&$-\tfrac{5}{6}$&$\tfrac{1}{6}$&$-\tfrac{1}{3}$&$\tfrac{1}{2}$&$-\tfrac{1}{6}$&$\tfrac{8}{9}$&$0$\\
$\bar{d}_{4}$&$2$&$0$&$0$&$0$&$1$&$-\tfrac{2}{3}$&${\color{red}-\tfrac{4}{3}\normalcolor}$&$0$&$\left(\mathbf{\bar{3}},\,\mathbf{1},\,\mathbf{1},\,\mathbf{1}\right)$&$-\tfrac{1}{3}$&$\tfrac{1}{6}$&$\tfrac{1}{2}$&$-\tfrac{1}{6}$&$\tfrac{1}{6}$&$-\tfrac{1}{3}$&$\tfrac{2}{3}$&$0$&$\tfrac{1}{3}$&$\tfrac{8}{9}$&$\tfrac{1}{3}$\\
\hline
$\delta_{5}$&$2$&$0$&$0$&$0$&$1$&$-\tfrac{2}{3}$&$-\tfrac{1}{3}$&$0$&$\left(\mathbf{3},\,\mathbf{1},\,\mathbf{1},\,\mathbf{1}\right)$&$\tfrac{1}{3}$&$-\tfrac{1}{3}$&$0$&$0$&$-1$&$\tfrac{2}{3}$&$0$&$0$&$0$&$-\tfrac{7}{9}$&$\tfrac{2}{3}$\\
$h_{10}$&$2$&$0$&$0$&$0$&$1$&$-\tfrac{2}{3}$&$-\tfrac{1}{3}$&$0$&$\left(\mathbf{1},\,\mathbf{1},\,\mathbf{1},\,\mathbf{2}\right)$&$0$&$\tfrac{2}{3}$&$0$&$0$&$0$&$-\tfrac{1}{3}$&$-1$&$0$&$0$&$\tfrac{2}{9}$&$0$\\
$\bar{\delta}_{5}$&$2$&$0$&$0$&$0$&$1$&$-\tfrac{2}{3}$&$-\tfrac{1}{3}$&$0$&$\left(\mathbf{\bar{3}},\,\mathbf{1},\,\mathbf{1},\,\mathbf{1}\right)$&$-\tfrac{1}{3}$&$-\tfrac{1}{3}$&$0$&$0$&$1$&$\tfrac{2}{3}$&$0$&$0$&$0$&$-\tfrac{1}{9}$&$-\tfrac{2}{3}$\\
$h_{9}$&$2$&$0$&$0$&$0$&$1$&$-\tfrac{2}{3}$&$-\tfrac{1}{3}$&$0$&$\left(\mathbf{1},\,\mathbf{1},\,\mathbf{1},\,\mathbf{2}\right)$&$0$&$\tfrac{2}{3}$&$0$&$0$&$0$&$-\tfrac{1}{3}$&$1$&$0$&$0$&$\tfrac{8}{9}$&$0$\\
$s^0_{28}$&$2$&$0$&$0$&$0$&$1$&$-\tfrac{2}{3}$&$-\tfrac{1}{3}$&$0$&$\left(\mathbf{1},\,\mathbf{1},\,\mathbf{1},\,\mathbf{1}\right)$&$0$&$\tfrac{2}{3}$&$0$&$0$&$0$&$-\tfrac{1}{3}$&$0$&$-1$&$0$&$-\tfrac{13}{9}$&$0$\\
$s^0_{27}$&$2$&$0$&$0$&$0$&$1$&$-\tfrac{2}{3}$&$-\tfrac{1}{3}$&$0$&$\left(\mathbf{1},\,\mathbf{1},\,\mathbf{1},\,\mathbf{1}\right)$&$0$&$\tfrac{2}{3}$&$0$&$0$&$0$&$-\tfrac{1}{3}$&$0$&$1$&$0$&$\tfrac{23}{9}$&$0$\\
$s^0_{29}$&$2$&$0$&$0$&$0$&$1$&$-\tfrac{2}{3}$&$\tfrac{2}{3}$&$0$&$\left(\mathbf{1},\,\mathbf{1},\,\mathbf{1},\,\mathbf{1}\right)$&$0$&$\tfrac{2}{3}$&$0$&$0$&$0$&$\tfrac{2}{3}$&$0$&$0$&$0$&$\tfrac{2}{9}$&$0$\\
\hline
$\bar{n}_{14}$&$2$&$2$&$0$&$0$&$1$&$-\tfrac{2}{3}$&$-\tfrac{1}{3}$&$0$&$\left(\mathbf{1},\,\mathbf{1},\,\mathbf{1},\,\mathbf{1}\right)$&$0$&$\tfrac{1}{6}$&$-\tfrac{1}{2}$&$\tfrac{1}{6}$&$\tfrac{5}{6}$&$-\tfrac{1}{3}$&$\tfrac{4}{3}$&$0$&$-\tfrac{1}{3}$&$\tfrac{5}{9}$&$-1$\\
$\bar{n}_{13}$&$2$&$2$&$0$&$0$&$1$&$-\tfrac{2}{3}$&$-\tfrac{1}{3}$&$0$&$\left(\mathbf{1},\,\mathbf{1},\,\mathbf{1},\,\mathbf{1}\right)$&$0$&$\tfrac{1}{6}$&$\tfrac{1}{2}$&$-\tfrac{5}{6}$&$\tfrac{5}{6}$&$-\tfrac{1}{3}$&$-\tfrac{2}{3}$&$0$&$-\tfrac{1}{3}$&$\tfrac{14}{9}$&$-1$\\
$\bar{\eta}_{5}$&$2$&$2$&$0$&$0$&$1$&$-\tfrac{2}{3}$&${\color{red}\tfrac{2}{3}\normalcolor}$&$0$&$\left(\mathbf{1},\,\mathbf{1},\,\mathbf{1},\,\mathbf{2}\right)$&$0$&$\tfrac{1}{6}$&$-\tfrac{1}{2}$&$\tfrac{1}{6}$&$\tfrac{5}{6}$&$\tfrac{2}{3}$&$\tfrac{1}{3}$&$0$&$-\tfrac{1}{3}$&$-\tfrac{1}{9}$&$-1$\\
$\bar{n}_{15}$&$2$&$2$&$0$&$0$&$1$&$-\tfrac{2}{3}$&${\color{red}-\tfrac{1}{3}\normalcolor}$&$0$&$\left(\mathbf{1},\,\mathbf{1},\,\mathbf{1},\,\mathbf{1}\right)$&$0$&$\tfrac{1}{6}$&$-\tfrac{1}{2}$&$\tfrac{1}{6}$&$\tfrac{5}{6}$&$-\tfrac{1}{3}$&$-\tfrac{2}{3}$&$0$&$-\tfrac{1}{3}$&$-\tfrac{1}{9}$&$-1$\\
\hline
$s^+_{14}$&$3$&$0$&$1$&$1$&$\tfrac{1}{3}$&${\color{red}-\tfrac{5}{2}\normalcolor}$&$0$&$-\tfrac{1}{2}$&$\left(\mathbf{1},\,\mathbf{1},\,\mathbf{1},\,\mathbf{1}\right)$&$-\tfrac{1}{2}$&$0$&$0$&$0$&$-1$&$-\tfrac{1}{2}$&$1$&$-\tfrac{1}{2}$&$0$&$-\tfrac{5}{6}$&$0$\\
$s^-_{14}$&$3$&$0$&$1$&$1$&$\tfrac{1}{3}$&${\color{red}-\tfrac{5}{2}\normalcolor}$&$0$&$-\tfrac{1}{2}$&$\left(\mathbf{1},\,\mathbf{1},\,\mathbf{1},\,\mathbf{1}\right)$&$\tfrac{1}{2}$&$0$&$0$&$0$&$1$&$-\tfrac{1}{2}$&$-1$&$-\tfrac{1}{2}$&$0$&$-\tfrac{5}{6}$&$0$\\
$s^+_{12}$&$3$&$0$&$1$&$1$&$0$&$-\tfrac{1}{2}$&$0$&$-\tfrac{1}{2}$&$\left(\mathbf{1},\,\mathbf{1},\,\mathbf{1},\,\mathbf{1}\right)$&$-\tfrac{1}{2}$&$0$&$0$&$0$&$-1$&$\tfrac{1}{2}$&$1$&$\tfrac{1}{2}$&$0$&$\tfrac{5}{6}$&$0$\\
$s^-_{12}$&$3$&$0$&$1$&$1$&$0$&$-\tfrac{1}{2}$&$0$&$-\tfrac{1}{2}$&$\left(\mathbf{1},\,\mathbf{1},\,\mathbf{1},\,\mathbf{1}\right)$&$\tfrac{1}{2}$&$0$&$0$&$0$&$1$&$\tfrac{1}{2}$&$-1$&$\tfrac{1}{2}$&$0$&$\tfrac{5}{6}$&$0$\\
$\bar{f}^+_{2}$&$3$&$0$&$1$&$1$&$-\tfrac{1}{3}$&${\color{red}\tfrac{3}{2}\normalcolor}$&$0$&${\color{red}\tfrac{1}{2}\normalcolor}$&$\left(\mathbf{1},\,\mathbf{1},\,\mathbf{\bar{4}},\,\mathbf{1}\right)$&$-\tfrac{1}{2}$&$0$&$0$&$0$&$-1$&$0$&$0$&$0$&$-\tfrac{1}{2}$&$-\tfrac{1}{2}$&$-1$\\
$\bar{f}^-_{2}$&$3$&$0$&$1$&$1$&$-\tfrac{1}{3}$&${\color{red}\tfrac{3}{2}\normalcolor}$&$0$&${\color{red}\tfrac{1}{2}\normalcolor}$&$\left(\mathbf{1},\,\mathbf{1},\,\mathbf{4},\,\mathbf{1}\right)$&$\tfrac{1}{2}$&$0$&$0$&$0$&$1$&$0$&$0$&$0$&$\tfrac{1}{2}$&$\tfrac{1}{2}$&$1$\\
\hline
$s^+_{11}$&$3$&$0$&$1$&$0$&$\tfrac{1}{3}$&${\color{red}-\tfrac{5}{2}\normalcolor}$&$0$&$-\tfrac{1}{2}$&$\left(\mathbf{1},\,\mathbf{1},\,\mathbf{1},\,\mathbf{1}\right)$&$-\tfrac{1}{2}$&$0$&$0$&$0$&$-1$&$-\tfrac{1}{2}$&$1$&$-\tfrac{1}{2}$&$0$&$-\tfrac{5}{6}$&$0$\\
$s^-_{11}$&$3$&$0$&$1$&$0$&$\tfrac{1}{3}$&${\color{red}-\tfrac{5}{2}\normalcolor}$&$0$&$-\tfrac{1}{2}$&$\left(\mathbf{1},\,\mathbf{1},\,\mathbf{1},\,\mathbf{1}\right)$&$\tfrac{1}{2}$&$0$&$0$&$0$&$1$&$-\tfrac{1}{2}$&$-1$&$-\tfrac{1}{2}$&$0$&$-\tfrac{5}{6}$&$0$\\
$s^+_{9}$&$3$&$0$&$1$&$0$&$0$&$-\tfrac{1}{2}$&$0$&$-\tfrac{1}{2}$&$\left(\mathbf{1},\,\mathbf{1},\,\mathbf{1},\,\mathbf{1}\right)$&$-\tfrac{1}{2}$&$0$&$0$&$0$&$-1$&$\tfrac{1}{2}$&$1$&$\tfrac{1}{2}$&$0$&$\tfrac{5}{6}$&$0$\\
$s^-_{9}$&$3$&$0$&$1$&$0$&$0$&$-\tfrac{1}{2}$&$0$&$-\tfrac{1}{2}$&$\left(\mathbf{1},\,\mathbf{1},\,\mathbf{1},\,\mathbf{1}\right)$&$\tfrac{1}{2}$&$0$&$0$&$0$&$1$&$\tfrac{1}{2}$&$-1$&$\tfrac{1}{2}$&$0$&$\tfrac{5}{6}$&$0$\\
$\bar{f}^+_{1}$&$3$&$0$&$1$&$0$&$-\tfrac{1}{3}$&${\color{red}\tfrac{3}{2}\normalcolor}$&$0$&${\color{red}\tfrac{1}{2}\normalcolor}$&$\left(\mathbf{1},\,\mathbf{1},\,\mathbf{\bar{4}},\,\mathbf{1}\right)$&$-\tfrac{1}{2}$&$0$&$0$&$0$&$-1$&$0$&$0$&$0$&$-\tfrac{1}{2}$&$-\tfrac{1}{2}$&$-1$\\
$\bar{f}^-_{1}$&$3$&$0$&$1$&$0$&$-\tfrac{1}{3}$&${\color{red}\tfrac{3}{2}\normalcolor}$&$0$&${\color{red}\tfrac{1}{2}\normalcolor}$&$\left(\mathbf{1},\,\mathbf{1},\,\mathbf{4},\,\mathbf{1}\right)$&$\tfrac{1}{2}$&$0$&$0$&$0$&$1$&$0$&$0$&$0$&$\tfrac{1}{2}$&$\tfrac{1}{2}$&$1$\\
\hline
$h_{6}$&$3$&$0$&$0$&$1$&$-\tfrac{1}{3}$&${\color{red}\tfrac{3}{2}\normalcolor}$&$0$&$-\tfrac{1}{2}$&$\left(\mathbf{1},\,\mathbf{1},\,\mathbf{1},\,\mathbf{2}\right)$&$0$&$0$&$-\tfrac{1}{2}$&$\tfrac{1}{2}$&$0$&$0$&$1$&$0$&$0$&$-\tfrac{1}{2}$&$0$\\
$h_{5}$&$3$&$0$&$0$&$1$&$-\tfrac{1}{3}$&${\color{red}\tfrac{3}{2}\normalcolor}$&$0$&$-\tfrac{1}{2}$&$\left(\mathbf{1},\,\mathbf{1},\,\mathbf{1},\,\mathbf{2}\right)$&$0$&$0$&$\tfrac{1}{2}$&$-\tfrac{1}{2}$&$0$&$0$&$-1$&$0$&$0$&$\tfrac{1}{2}$&$0$\\
$\chi_{4}$&$3$&$0$&$0$&$1$&$-\tfrac{1}{3}$&${\color{red}\tfrac{3}{2}\normalcolor}$&$0$&$-\tfrac{1}{2}$&$\left(\mathbf{1},\,\mathbf{1},\,\mathbf{1},\,\mathbf{1}\right)$&$0$&$0$&$-\tfrac{1}{2}$&$\tfrac{1}{2}$&$0$&$0$&$0$&$0$&$1$&$-\tfrac{1}{2}$&$2$\\
$\chi_{3}$&$3$&$0$&$0$&$1$&$-\tfrac{1}{3}$&${\color{red}\tfrac{3}{2}\normalcolor}$&$0$&$-\tfrac{1}{2}$&$\left(\mathbf{1},\,\mathbf{1},\,\mathbf{1},\,\mathbf{1}\right)$&$0$&$0$&$\tfrac{1}{2}$&$-\tfrac{1}{2}$&$0$&$0$&$0$&$0$&$-1$&$\tfrac{1}{2}$&$-2$\\
\hline
$h_{4}$&$3$&$0$&$0$&$0$&$-\tfrac{1}{3}$&${\color{red}\tfrac{3}{2}\normalcolor}$&$0$&$-\tfrac{1}{2}$&$\left(\mathbf{1},\,\mathbf{1},\,\mathbf{1},\,\mathbf{2}\right)$&$0$&$0$&$-\tfrac{1}{2}$&$\tfrac{1}{2}$&$0$&$0$&$1$&$0$&$0$&$-\tfrac{1}{2}$&$0$\\
$h_{3}$&$3$&$0$&$0$&$0$&$-\tfrac{1}{3}$&${\color{red}\tfrac{3}{2}\normalcolor}$&$0$&$-\tfrac{1}{2}$&$\left(\mathbf{1},\,\mathbf{1},\,\mathbf{1},\,\mathbf{2}\right)$&$0$&$0$&$\tfrac{1}{2}$&$-\tfrac{1}{2}$&$0$&$0$&$-1$&$0$&$0$&$\tfrac{1}{2}$&$0$\\
$\chi_{2}$&$3$&$0$&$0$&$0$&$-\tfrac{1}{3}$&${\color{red}\tfrac{3}{2}\normalcolor}$&$0$&$-\tfrac{1}{2}$&$\left(\mathbf{1},\,\mathbf{1},\,\mathbf{1},\,\mathbf{1}\right)$&$0$&$0$&$-\tfrac{1}{2}$&$\tfrac{1}{2}$&$0$&$0$&$0$&$0$&$1$&$-\tfrac{1}{2}$&$2$\\
$\chi_{1}$&$3$&$0$&$0$&$0$&$-\tfrac{1}{3}$&${\color{red}\tfrac{3}{2}\normalcolor}$&$0$&$-\tfrac{1}{2}$&$\left(\mathbf{1},\,\mathbf{1},\,\mathbf{1},\,\mathbf{1}\right)$&$0$&$0$&$\tfrac{1}{2}$&$-\tfrac{1}{2}$&$0$&$0$&$0$&$0$&$-1$&$\tfrac{1}{2}$&$-2$\\
\hline
$s^+_{13}$&$3$&$0$&$1$&$1$&$1$&$-\tfrac{1}{2}$&$0$&$-\tfrac{1}{2}$&$\left(\mathbf{1},\,\mathbf{1},\,\mathbf{1},\,\mathbf{1}\right)$&$-\tfrac{1}{2}$&$0$&$0$&$0$&$-1$&$\tfrac{1}{2}$&$1$&$\tfrac{1}{2}$&$0$&$\tfrac{5}{6}$&$0$\\
$s^-_{13}$&$3$&$0$&$1$&$1$&$1$&$-\tfrac{1}{2}$&$0$&$-\tfrac{1}{2}$&$\left(\mathbf{1},\,\mathbf{1},\,\mathbf{1},\,\mathbf{1}\right)$&$\tfrac{1}{2}$&$0$&$0$&$0$&$1$&$\tfrac{1}{2}$&$-1$&$\tfrac{1}{2}$&$0$&$\tfrac{5}{6}$&$0$\\
\hline
$s^+_{10}$&$3$&$0$&$1$&$0$&$1$&$-\tfrac{1}{2}$&$0$&$-\tfrac{1}{2}$&$\left(\mathbf{1},\,\mathbf{1},\,\mathbf{1},\,\mathbf{1}\right)$&$-\tfrac{1}{2}$&$0$&$0$&$0$&$-1$&$\tfrac{1}{2}$&$1$&$\tfrac{1}{2}$&$0$&$\tfrac{5}{6}$&$0$\\
$s^-_{10}$&$3$&$0$&$1$&$0$&$1$&$-\tfrac{1}{2}$&$0$&$-\tfrac{1}{2}$&$\left(\mathbf{1},\,\mathbf{1},\,\mathbf{1},\,\mathbf{1}\right)$&$\tfrac{1}{2}$&$0$&$0$&$0$&$1$&$\tfrac{1}{2}$&$-1$&$\tfrac{1}{2}$&$0$&$\tfrac{5}{6}$&$0$\\
\hline
$s^0_{22}$&$4$&$2$&$0$&$0$&$0$&$-\tfrac{1}{3}$&${\color{red}\tfrac{1}{3}\normalcolor}$&$0$&$\left(\mathbf{1},\,\mathbf{1},\,\mathbf{1},\,\mathbf{1}\right)$&$0$&$-\tfrac{2}{3}$&$0$&$\tfrac{1}{3}$&$\tfrac{5}{3}$&$\tfrac{1}{3}$&$\tfrac{2}{3}$&$0$&$\tfrac{1}{3}$&$-\tfrac{2}{9}$&$0$\\
$f_{5}$&$4$&$2$&$0$&$0$&$0$&$-\tfrac{1}{3}$&${\color{red}\tfrac{1}{3}\normalcolor}$&$0$&$\left(\mathbf{1},\,\mathbf{1},\,\mathbf{4},\,\mathbf{1}\right)$&$0$&$-\tfrac{1}{6}$&$\tfrac{1}{2}$&$-\tfrac{1}{6}$&$-\tfrac{5}{6}$&$-\tfrac{1}{6}$&$-\tfrac{1}{3}$&$\tfrac{1}{2}$&$-\tfrac{1}{6}$&$\tfrac{7}{9}$&$0$\\
$l_{3}$&$4$&$2$&$0$&$0$&$0$&$-\tfrac{1}{3}$&$-\tfrac{2}{3}$&$0$&$\left(\mathbf{1},\,\mathbf{2},\,\mathbf{1},\,\mathbf{1}\right)$&$\tfrac{1}{2}$&$-\tfrac{1}{6}$&$-\tfrac{1}{2}$&$-\tfrac{1}{6}$&$\tfrac{1}{6}$&$\tfrac{1}{3}$&$\tfrac{2}{3}$&$0$&$\tfrac{1}{3}$&$\tfrac{4}{9}$&$1$\\
$n_{10}$&$4$&$2$&$0$&$0$&$\tfrac{1}{2}$&${\color{red}\tfrac{8}{3}\normalcolor}$&$-\tfrac{2}{3}$&$0$&$\left(\mathbf{1},\,\mathbf{1},\,\mathbf{1},\,\mathbf{1}\right)$&$0$&$-\tfrac{1}{6}$&$-\tfrac{1}{2}$&$\tfrac{5}{6}$&$-\tfrac{5}{6}$&$\tfrac{1}{3}$&$\tfrac{2}{3}$&$0$&$\tfrac{1}{3}$&$-\tfrac{14}{9}$&$1$\\
$n_{9}$&$4$&$2$&$0$&$0$&$\tfrac{1}{2}$&${\color{red}\tfrac{8}{3}\normalcolor}$&$-\tfrac{2}{3}$&$0$&$\left(\mathbf{1},\,\mathbf{1},\,\mathbf{1},\,\mathbf{1}\right)$&$0$&$-\tfrac{1}{6}$&$\tfrac{1}{2}$&$-\tfrac{1}{6}$&$-\tfrac{5}{6}$&$\tfrac{1}{3}$&$-\tfrac{4}{3}$&$0$&$\tfrac{1}{3}$&$-\tfrac{5}{9}$&$1$\\
$\eta_{5}$&$4$&$2$&$0$&$0$&$\tfrac{1}{2}$&${\color{red}\tfrac{8}{3}\normalcolor}$&${\color{red}\tfrac{4}{3}\normalcolor}$&$0$&$\left(\mathbf{1},\,\mathbf{1},\,\mathbf{1},\,\mathbf{2}\right)$&$0$&$-\tfrac{1}{6}$&$\tfrac{1}{2}$&$-\tfrac{1}{6}$&$-\tfrac{5}{6}$&$-\tfrac{2}{3}$&$-\tfrac{1}{3}$&$0$&$\tfrac{1}{3}$&$\tfrac{1}{9}$&$1$\\
$\delta_{2}$&$4$&$2$&$0$&$0$&$0$&$-\tfrac{1}{3}$&${\color{red}\tfrac{4}{3}\normalcolor}$&$0$&$\left(\mathbf{3},\,\mathbf{1},\,\mathbf{1},\,\mathbf{1}\right)$&$\tfrac{1}{3}$&$\tfrac{1}{3}$&$0$&$\tfrac{1}{3}$&$\tfrac{2}{3}$&$\tfrac{1}{3}$&$\tfrac{2}{3}$&$0$&$\tfrac{1}{3}$&$\tfrac{1}{9}$&$\tfrac{2}{3}$\\
$n_{7}$&$4$&$2$&$0$&$0$&$0$&$\tfrac{2}{3}$&${\color{red}\tfrac{1}{3}\normalcolor}$&$0$&$\left(\mathbf{1},\,\mathbf{1},\,\mathbf{1},\,\mathbf{1}\right)$&$0$&$-\tfrac{1}{6}$&$\tfrac{1}{2}$&$-\tfrac{1}{6}$&$-\tfrac{5}{6}$&$\tfrac{1}{3}$&$\tfrac{2}{3}$&$0$&$\tfrac{1}{3}$&$\tfrac{1}{9}$&$1$\\
$n_{11}$&$4$&$2$&$0$&$0$&$\tfrac{1}{2}$&${\color{red}\tfrac{8}{3}\normalcolor}$&${\color{red}-\tfrac{2}{3}\normalcolor}$&$0$&$\left(\mathbf{1},\,\mathbf{1},\,\mathbf{1},\,\mathbf{1}\right)$&$0$&$-\tfrac{1}{6}$&$\tfrac{1}{2}$&$-\tfrac{1}{6}$&$-\tfrac{5}{6}$&$\tfrac{1}{3}$&$\tfrac{2}{3}$&$0$&$\tfrac{1}{3}$&$\tfrac{1}{9}$&$1$\\
\hline
$h_{1}$&$4$&$0$&$0$&$0$&$\tfrac{1}{2}$&${\color{red}\tfrac{8}{3}\normalcolor}$&$-\tfrac{2}{3}$&$0$&$\left(\mathbf{1},\,\mathbf{1},\,\mathbf{1},\,\mathbf{2}\right)$&$0$&$-\tfrac{2}{3}$&$0$&$0$&$0$&$\tfrac{1}{3}$&$1$&$0$&$0$&$-\tfrac{2}{9}$&$0$\\
$\bar{\delta}_{1}$&$4$&$0$&$0$&$0$&$\tfrac{1}{2}$&${\color{red}\tfrac{8}{3}\normalcolor}$&$-\tfrac{2}{3}$&$0$&$\left(\mathbf{\bar{3}},\,\mathbf{1},\,\mathbf{1},\,\mathbf{1}\right)$&$-\tfrac{1}{3}$&$\tfrac{1}{3}$&$0$&$0$&$1$&$-\tfrac{2}{3}$&$0$&$0$&$0$&$\tfrac{7}{9}$&$-\tfrac{2}{3}$\\
$h_{2}$&$4$&$0$&$0$&$0$&$\tfrac{1}{2}$&${\color{red}\tfrac{8}{3}\normalcolor}$&$-\tfrac{2}{3}$&$0$&$\left(\mathbf{1},\,\mathbf{1},\,\mathbf{1},\,\mathbf{2}\right)$&$0$&$-\tfrac{2}{3}$&$0$&$0$&$0$&$\tfrac{1}{3}$&$-1$&$0$&$0$&$-\tfrac{8}{9}$&$0$\\
$\delta_{1}$&$4$&$0$&$0$&$0$&$\tfrac{1}{2}$&${\color{red}\tfrac{8}{3}\normalcolor}$&$-\tfrac{2}{3}$&$0$&$\left(\mathbf{3},\,\mathbf{1},\,\mathbf{1},\,\mathbf{1}\right)$&$\tfrac{1}{3}$&$\tfrac{1}{3}$&$0$&$0$&$-1$&$-\tfrac{2}{3}$&$0$&$0$&$0$&$\tfrac{1}{9}$&$\tfrac{2}{3}$\\
$s^0_{18}$&$4$&$0$&$0$&$0$&$\tfrac{1}{2}$&${\color{red}\tfrac{8}{3}\normalcolor}$&$-\tfrac{2}{3}$&$0$&$\left(\mathbf{1},\,\mathbf{1},\,\mathbf{1},\,\mathbf{1}\right)$&$0$&$-\tfrac{2}{3}$&$0$&$0$&$0$&$\tfrac{1}{3}$&$0$&$-1$&$0$&$-\tfrac{23}{9}$&$0$\\
$s^0_{17}$&$4$&$0$&$0$&$0$&$\tfrac{1}{2}$&${\color{red}\tfrac{8}{3}\normalcolor}$&$-\tfrac{2}{3}$&$0$&$\left(\mathbf{1},\,\mathbf{1},\,\mathbf{1},\,\mathbf{1}\right)$&$0$&$-\tfrac{2}{3}$&$0$&$0$&$0$&$\tfrac{1}{3}$&$0$&$1$&$0$&$\tfrac{13}{9}$&$0$\\
$s^0_{15}$&$4$&$0$&$0$&$0$&$0$&$\tfrac{2}{3}$&$-\tfrac{2}{3}$&$0$&$\left(\mathbf{1},\,\mathbf{1},\,\mathbf{1},\,\mathbf{1}\right)$&$0$&$-\tfrac{2}{3}$&$0$&$0$&$0$&$-\tfrac{2}{3}$&$0$&$0$&$0$&$-\tfrac{2}{9}$&$0$\\
$s^0_{19}$&$4$&$0$&$0$&$0$&$\tfrac{1}{2}$&${\color{red}\tfrac{8}{3}\normalcolor}$&$\tfrac{4}{3}$&$0$&$\left(\mathbf{1},\,\mathbf{1},\,\mathbf{1},\,\mathbf{1}\right)$&$0$&$-\tfrac{2}{3}$&$0$&$0$&$0$&$-\tfrac{2}{3}$&$0$&$0$&$0$&$-\tfrac{2}{9}$&$0$\\
\hline
$s^0_{20}$&$4$&$1$&$0$&$0$&$0$&$-\tfrac{1}{3}$&${\color{red}\tfrac{4}{3}\normalcolor}$&$0$&$\left(\mathbf{1},\,\mathbf{1},\,\mathbf{1},\,\mathbf{1}\right)$&$0$&$-\tfrac{2}{3}$&$0$&$-\tfrac{1}{3}$&$-\tfrac{5}{3}$&$\tfrac{1}{3}$&$-\tfrac{2}{3}$&$0$&$-\tfrac{1}{3}$&$-\tfrac{8}{9}$&$0$\\
$d_{1}$&$4$&$1$&$0$&$0$&$\tfrac{1}{2}$&${\color{red}\tfrac{8}{3}\normalcolor}$&${\color{red}\tfrac{1}{3}\normalcolor}$&$0$&$\left(\mathbf{3},\,\mathbf{1},\,\mathbf{1},\,\mathbf{1}\right)$&$\tfrac{1}{3}$&$-\tfrac{1}{6}$&$-\tfrac{1}{2}$&$\tfrac{1}{6}$&$-\tfrac{1}{6}$&$\tfrac{1}{3}$&$-\tfrac{2}{3}$&$0$&$-\tfrac{1}{3}$&$-\tfrac{8}{9}$&$-\tfrac{1}{3}$\\
$\eta_{3}$&$4$&$1$&$0$&$0$&$0$&$-\tfrac{1}{3}$&${\color{red}\tfrac{1}{3}\normalcolor}$&$0$&$\left(\mathbf{1},\,\mathbf{1},\,\mathbf{1},\,\mathbf{2}\right)$&$0$&$-\tfrac{1}{6}$&$\tfrac{1}{2}$&$\tfrac{1}{6}$&$\tfrac{5}{6}$&$\tfrac{1}{3}$&$\tfrac{1}{3}$&$0$&$\tfrac{2}{3}$&$\tfrac{1}{9}$&$1$\\
$f_{4}$&$4$&$1$&$0$&$0$&$\tfrac{1}{2}$&${\color{red}\tfrac{8}{3}\normalcolor}$&${\color{red}\tfrac{4}{3}\normalcolor}$&$0$&$\left(\mathbf{1},\,\mathbf{1},\,\mathbf{4},\,\mathbf{1}\right)$&$0$&$-\tfrac{1}{6}$&$\tfrac{1}{2}$&$\tfrac{1}{6}$&$\tfrac{5}{6}$&$-\tfrac{1}{6}$&$\tfrac{1}{3}$&$-\tfrac{1}{2}$&$\tfrac{1}{6}$&$-\tfrac{8}{9}$&$0$\\
$\bar{n}_{8}$&$4$&$1$&$0$&$0$&$\tfrac{1}{2}$&${\color{red}\tfrac{8}{3}\normalcolor}$&${\color{red}\tfrac{4}{3}\normalcolor}$&$0$&$\left(\mathbf{1},\,\mathbf{1},\,\mathbf{1},\,\mathbf{1}\right)$&$0$&$\tfrac{5}{6}$&$-\tfrac{1}{2}$&$\tfrac{1}{6}$&$\tfrac{5}{6}$&$\tfrac{1}{3}$&$-\tfrac{2}{3}$&$0$&$-\tfrac{1}{3}$&$\tfrac{1}{9}$&$-1$\\
$n_{5}$&$4$&$1$&$0$&$0$&$0$&$-\tfrac{1}{3}$&$-\tfrac{2}{3}$&$0$&$\left(\mathbf{1},\,\mathbf{1},\,\mathbf{1},\,\mathbf{1}\right)$&$0$&$-\tfrac{1}{6}$&$\tfrac{1}{2}$&$\tfrac{1}{6}$&$\tfrac{5}{6}$&$-\tfrac{2}{3}$&$-\tfrac{2}{3}$&$0$&$\tfrac{2}{3}$&$\tfrac{1}{9}$&$1$\\
$\bar{\delta}_{2}$&$4$&$1$&$0$&$0$&$0$&$-\tfrac{1}{3}$&$-\tfrac{2}{3}$&$0$&$\left(\mathbf{\bar{3}},\,\mathbf{1},\,\mathbf{1},\,\mathbf{1}\right)$&$-\tfrac{1}{3}$&$\tfrac{1}{3}$&$0$&$-\tfrac{1}{3}$&$-\tfrac{2}{3}$&$\tfrac{1}{3}$&$-\tfrac{2}{3}$&$0$&$-\tfrac{1}{3}$&$\tfrac{1}{9}$&$-\tfrac{2}{3}$\\
\hline
$s^0_{23}$&$4$&$2$&$0$&$0$&$1$&$-\tfrac{1}{3}$&${\color{red}\tfrac{1}{3}\normalcolor}$&$0$&$\left(\mathbf{1},\,\mathbf{1},\,\mathbf{1},\,\mathbf{1}\right)$&$0$&$-\tfrac{2}{3}$&$0$&$\tfrac{1}{3}$&$\tfrac{5}{3}$&$\tfrac{1}{3}$&$\tfrac{2}{3}$&$0$&$\tfrac{1}{3}$&$-\tfrac{2}{9}$&$0$\\
$f_{6}$&$4$&$2$&$0$&$0$&$1$&$-\tfrac{1}{3}$&${\color{red}\tfrac{1}{3}\normalcolor}$&$0$&$\left(\mathbf{1},\,\mathbf{1},\,\mathbf{4},\,\mathbf{1}\right)$&$0$&$-\tfrac{1}{6}$&$\tfrac{1}{2}$&$-\tfrac{1}{6}$&$-\tfrac{5}{6}$&$-\tfrac{1}{6}$&$-\tfrac{1}{3}$&$\tfrac{1}{2}$&$-\tfrac{1}{6}$&$\tfrac{7}{9}$&$0$\\
$l_{4}$&$4$&$2$&$0$&$0$&$1$&$-\tfrac{1}{3}$&$-\tfrac{2}{3}$&$0$&$\left(\mathbf{1},\,\mathbf{2},\,\mathbf{1},\,\mathbf{1}\right)$&$\tfrac{1}{2}$&$-\tfrac{1}{6}$&$-\tfrac{1}{2}$&$-\tfrac{1}{6}$&$\tfrac{1}{6}$&$\tfrac{1}{3}$&$\tfrac{2}{3}$&$0$&$\tfrac{1}{3}$&$\tfrac{4}{9}$&$1$\\
$\delta_{3}$&$4$&$2$&$0$&$0$&$1$&$-\tfrac{1}{3}$&${\color{red}\tfrac{4}{3}\normalcolor}$&$0$&$\left(\mathbf{3},\,\mathbf{1},\,\mathbf{1},\,\mathbf{1}\right)$&$\tfrac{1}{3}$&$\tfrac{1}{3}$&$0$&$\tfrac{1}{3}$&$\tfrac{2}{3}$&$\tfrac{1}{3}$&$\tfrac{2}{3}$&$0$&$\tfrac{1}{3}$&$\tfrac{1}{9}$&$\tfrac{2}{3}$\\
$n_{8}$&$4$&$2$&$0$&$0$&$1$&$\tfrac{2}{3}$&${\color{red}\tfrac{1}{3}\normalcolor}$&$0$&$\left(\mathbf{1},\,\mathbf{1},\,\mathbf{1},\,\mathbf{1}\right)$&$0$&$-\tfrac{1}{6}$&$\tfrac{1}{2}$&$-\tfrac{1}{6}$&$-\tfrac{5}{6}$&$\tfrac{1}{3}$&$\tfrac{2}{3}$&$0$&$\tfrac{1}{3}$&$\tfrac{1}{9}$&$1$\\
\hline
$s^0_{16}$&$4$&$0$&$0$&$0$&$1$&$\tfrac{2}{3}$&$-\tfrac{2}{3}$&$0$&$\left(\mathbf{1},\,\mathbf{1},\,\mathbf{1},\,\mathbf{1}\right)$&$0$&$-\tfrac{2}{3}$&$0$&$0$&$0$&$-\tfrac{2}{3}$&$0$&$0$&$0$&$-\tfrac{2}{9}$&$0$\\
\hline
$s^0_{21}$&$4$&$1$&$0$&$0$&$1$&$-\tfrac{1}{3}$&${\color{red}\tfrac{4}{3}\normalcolor}$&$0$&$\left(\mathbf{1},\,\mathbf{1},\,\mathbf{1},\,\mathbf{1}\right)$&$0$&$-\tfrac{2}{3}$&$0$&$-\tfrac{1}{3}$&$-\tfrac{5}{3}$&$\tfrac{1}{3}$&$-\tfrac{2}{3}$&$0$&$-\tfrac{1}{3}$&$-\tfrac{8}{9}$&$0$\\
$\eta_{4}$&$4$&$1$&$0$&$0$&$1$&$-\tfrac{1}{3}$&${\color{red}\tfrac{1}{3}\normalcolor}$&$0$&$\left(\mathbf{1},\,\mathbf{1},\,\mathbf{1},\,\mathbf{2}\right)$&$0$&$-\tfrac{1}{6}$&$\tfrac{1}{2}$&$\tfrac{1}{6}$&$\tfrac{5}{6}$&$\tfrac{1}{3}$&$\tfrac{1}{3}$&$0$&$\tfrac{2}{3}$&$\tfrac{1}{9}$&$1$\\
$n_{6}$&$4$&$1$&$0$&$0$&$1$&$-\tfrac{1}{3}$&$-\tfrac{2}{3}$&$0$&$\left(\mathbf{1},\,\mathbf{1},\,\mathbf{1},\,\mathbf{1}\right)$&$0$&$-\tfrac{1}{6}$&$\tfrac{1}{2}$&$\tfrac{1}{6}$&$\tfrac{5}{6}$&$-\tfrac{2}{3}$&$-\tfrac{2}{3}$&$0$&$\tfrac{2}{3}$&$\tfrac{1}{9}$&$1$\\
$\bar{\delta}_{3}$&$4$&$1$&$0$&$0$&$1$&$-\tfrac{1}{3}$&$-\tfrac{2}{3}$&$0$&$\left(\mathbf{\bar{3}},\,\mathbf{1},\,\mathbf{1},\,\mathbf{1}\right)$&$-\tfrac{1}{3}$&$\tfrac{1}{3}$&$0$&$-\tfrac{1}{3}$&$-\tfrac{2}{3}$&$\tfrac{1}{3}$&$-\tfrac{2}{3}$&$0$&$-\tfrac{1}{3}$&$\tfrac{1}{9}$&$-\tfrac{2}{3}$\\
\hline
$s^+_{4}$&$5$&$1$&$1$&$1$&$0$&$-\tfrac{1}{6}$&${\color{red}\tfrac{2}{3}\normalcolor}$&${\color{red}\tfrac{1}{2}\normalcolor}$&$\left(\mathbf{1},\,\mathbf{1},\,\mathbf{1},\,\mathbf{1}\right)$&$-\tfrac{1}{2}$&$-\tfrac{1}{3}$&$0$&$-\tfrac{2}{3}$&$\tfrac{2}{3}$&$\tfrac{1}{6}$&$-\tfrac{1}{3}$&$\tfrac{1}{2}$&$\tfrac{1}{3}$&$\tfrac{37}{18}$&$0$\\
$\nu_{2}$&$5$&$1$&$1$&$1$&$0$&$-\tfrac{1}{6}$&${\color{red}\tfrac{2}{3}\normalcolor}$&$-\tfrac{1}{2}$&$\left(\mathbf{3},\,\mathbf{1},\,\mathbf{1},\,\mathbf{1}\right)$&$-\tfrac{1}{6}$&$-\tfrac{1}{3}$&$0$&$\tfrac{1}{3}$&$-\tfrac{1}{3}$&$\tfrac{1}{6}$&$-\tfrac{1}{3}$&$\tfrac{1}{2}$&$\tfrac{1}{3}$&$\tfrac{1}{18}$&$\tfrac{2}{3}$\\
$s^-_{3}$&$5$&$1$&$1$&$1$&$0$&$-\tfrac{1}{6}$&$-\tfrac{1}{3}$&$-\tfrac{1}{2}$&$\left(\mathbf{1},\,\mathbf{1},\,\mathbf{1},\,\mathbf{1}\right)$&$\tfrac{1}{2}$&$\tfrac{1}{6}$&$-\tfrac{1}{2}$&$-\tfrac{1}{6}$&$\tfrac{1}{6}$&$\tfrac{1}{6}$&$-\tfrac{1}{3}$&$-\tfrac{1}{2}$&$-\tfrac{2}{3}$&$-\tfrac{17}{18}$&$-1$\\
$m_{6}$&$5$&$1$&$1$&$1$&$0$&$-\tfrac{1}{6}$&$-\tfrac{1}{3}$&${\color{red}\tfrac{1}{2}\normalcolor}$&$\left(\mathbf{1},\,\mathbf{2},\,\mathbf{1},\,\mathbf{1}\right)$&$0$&$\tfrac{1}{6}$&$\tfrac{1}{2}$&$-\tfrac{1}{6}$&$-\tfrac{5}{6}$&$\tfrac{1}{6}$&$-\tfrac{1}{3}$&$\tfrac{1}{2}$&$\tfrac{1}{3}$&$\tfrac{19}{18}$&$1$\\
$s^+_{3}$&$5$&$1$&$1$&$1$&$0$&$-\tfrac{1}{6}$&${\color{red}-\tfrac{4}{3}\normalcolor}$&$-\tfrac{1}{2}$&$\left(\mathbf{1},\,\mathbf{1},\,\mathbf{1},\,\mathbf{1}\right)$&$-\tfrac{1}{2}$&$\tfrac{2}{3}$&$0$&$\tfrac{1}{3}$&$\tfrac{2}{3}$&$\tfrac{1}{6}$&$-\tfrac{1}{3}$&$\tfrac{1}{2}$&$\tfrac{1}{3}$&$\tfrac{19}{18}$&$0$\\
$s^-_{4}$&$5$&$1$&$1$&$1$&$0$&$\tfrac{5}{6}$&${\color{red}\tfrac{2}{3}\normalcolor}$&${\color{red}\tfrac{1}{2}\normalcolor}$&$\left(\mathbf{1},\,\mathbf{1},\,\mathbf{1},\,\mathbf{1}\right)$&$\tfrac{1}{2}$&$\tfrac{1}{6}$&$-\tfrac{1}{2}$&$-\tfrac{1}{6}$&$\tfrac{1}{6}$&$\tfrac{1}{6}$&$-\tfrac{1}{3}$&$\tfrac{1}{2}$&$\tfrac{1}{3}$&$\tfrac{25}{18}$&$1$\\
\hline
$s^+_{2}$&$5$&$1$&$1$&$0$&$0$&$-\tfrac{1}{6}$&${\color{red}\tfrac{2}{3}\normalcolor}$&${\color{red}\tfrac{1}{2}\normalcolor}$&$\left(\mathbf{1},\,\mathbf{1},\,\mathbf{1},\,\mathbf{1}\right)$&$-\tfrac{1}{2}$&$-\tfrac{1}{3}$&$0$&$-\tfrac{2}{3}$&$\tfrac{2}{3}$&$\tfrac{1}{6}$&$-\tfrac{1}{3}$&$\tfrac{1}{2}$&$\tfrac{1}{3}$&$\tfrac{37}{18}$&$0$\\
$nu_{1}$&$5$&$1$&$1$&$0$&$0$&$-\tfrac{1}{6}$&${\color{red}\tfrac{2}{3}\normalcolor}$&$-\tfrac{1}{2}$&$\left(\mathbf{3},\,\mathbf{1},\,\mathbf{1},\,\mathbf{1}\right)$&$-\tfrac{1}{6}$&$-\tfrac{1}{3}$&$0$&$\tfrac{1}{3}$&$-\tfrac{1}{3}$&$\tfrac{1}{6}$&$-\tfrac{1}{3}$&$\tfrac{1}{2}$&$\tfrac{1}{3}$&$\tfrac{1}{18}$&$\tfrac{2}{3}$\\
$s^-_{1}$&$5$&$1$&$1$&$0$&$0$&$-\tfrac{1}{6}$&$-\tfrac{1}{3}$&$-\tfrac{1}{2}$&$\left(\mathbf{1},\,\mathbf{1},\,\mathbf{1},\,\mathbf{1}\right)$&$\tfrac{1}{2}$&$\tfrac{1}{6}$&$-\tfrac{1}{2}$&$-\tfrac{1}{6}$&$\tfrac{1}{6}$&$\tfrac{1}{6}$&$-\tfrac{1}{3}$&$-\tfrac{1}{2}$&$-\tfrac{2}{3}$&$-\tfrac{17}{18}$&$-1$\\
$m_{5}$&$5$&$1$&$1$&$0$&$0$&$-\tfrac{1}{6}$&$-\tfrac{1}{3}$&${\color{red}\tfrac{1}{2}\normalcolor}$&$\left(\mathbf{1},\,\mathbf{2},\,\mathbf{1},\,\mathbf{1}\right)$&$0$&$\tfrac{1}{6}$&$\tfrac{1}{2}$&$-\tfrac{1}{6}$&$-\tfrac{5}{6}$&$\tfrac{1}{6}$&$-\tfrac{1}{3}$&$\tfrac{1}{2}$&$\tfrac{1}{3}$&$\tfrac{19}{18}$&$1$\\
$s^+_{1}$&$5$&$1$&$1$&$0$&$0$&$-\tfrac{1}{6}$&${\color{red}-\tfrac{4}{3}\normalcolor}$&$-\tfrac{1}{2}$&$\left(\mathbf{1},\,\mathbf{1},\,\mathbf{1},\,\mathbf{1}\right)$&$-\tfrac{1}{2}$&$\tfrac{2}{3}$&$0$&$\tfrac{1}{3}$&$\tfrac{2}{3}$&$\tfrac{1}{6}$&$-\tfrac{1}{3}$&$\tfrac{1}{2}$&$\tfrac{1}{3}$&$\tfrac{19}{18}$&$0$\\
$s^-_{2}$&$5$&$1$&$1$&$0$&$0$&$\tfrac{5}{6}$&${\color{red}\tfrac{2}{3}\normalcolor}$&${\color{red}\tfrac{1}{2}\normalcolor}$&$\left(\mathbf{1},\,\mathbf{1},\,\mathbf{1},\,\mathbf{1}\right)$&$\tfrac{1}{2}$&$\tfrac{1}{6}$&$-\tfrac{1}{2}$&$-\tfrac{1}{6}$&$\tfrac{1}{6}$&$\tfrac{1}{6}$&$-\tfrac{1}{3}$&$\tfrac{1}{2}$&$\tfrac{1}{3}$&$\tfrac{25}{18}$&$1$\\
\hline
$\eta_{2}$&$5$&$1$&$0$&$1$&$0$&$-\tfrac{1}{6}$&${\color{red}\tfrac{2}{3}\normalcolor}$&$-\tfrac{1}{2}$&$\left(\mathbf{1},\,\mathbf{1},\,\mathbf{1},\,\mathbf{2}\right)$&$0$&$\tfrac{1}{6}$&$0$&$-\tfrac{2}{3}$&$-\tfrac{5}{6}$&$-\tfrac{1}{3}$&$-\tfrac{1}{3}$&$0$&$\tfrac{1}{3}$&$\tfrac{19}{18}$&$1$\\
$\bar{n}_{5}$&$5$&$1$&$0$&$1$&$0$&$-\tfrac{1}{6}$&${\color{red}\tfrac{2}{3}\normalcolor}$&$-\tfrac{1}{2}$&$\left(\mathbf{1},\,\mathbf{1},\,\mathbf{1},\,\mathbf{1}\right)$&$0$&$\tfrac{1}{6}$&$0$&$-\tfrac{2}{3}$&$-\tfrac{5}{6}$&$-\tfrac{1}{3}$&$\tfrac{2}{3}$&$0$&$-\tfrac{2}{3}$&$\tfrac{19}{18}$&$-1$\\
$n_{2}$&$5$&$1$&$0$&$1$&$0$&$-\tfrac{1}{6}$&${\color{red}-\tfrac{4}{3}\normalcolor}$&$-\tfrac{1}{2}$&$\left(\mathbf{1},\,\mathbf{1},\,\mathbf{1},\,\mathbf{1}\right)$&$0$&$\tfrac{1}{6}$&$0$&$-\tfrac{2}{3}$&$-\tfrac{5}{6}$&$\tfrac{2}{3}$&$\tfrac{2}{3}$&$0$&$\tfrac{1}{3}$&$\tfrac{19}{18}$&$1$\\
\hline
$\eta_{1}$&$5$&$1$&$0$&$0$&$0$&$-\tfrac{1}{6}$&${\color{red}\tfrac{2}{3}\normalcolor}$&$-\tfrac{1}{2}$&$\left(\mathbf{1},\,\mathbf{1},\,\mathbf{1},\,\mathbf{2}\right)$&$0$&$\tfrac{1}{6}$&$0$&$-\tfrac{2}{3}$&$-\tfrac{5}{6}$&$-\tfrac{1}{3}$&$-\tfrac{1}{3}$&$0$&$\tfrac{1}{3}$&$\tfrac{19}{18}$&$1$\\
$n_{1}$&$5$&$1$&$0$&$0$&$0$&$-\tfrac{1}{6}$&${\color{red}\tfrac{2}{3}\normalcolor}$&$-\tfrac{1}{2}$&$\left(\mathbf{1},\,\mathbf{1},\,\mathbf{1},\,\mathbf{1}\right)$&$0$&$\tfrac{1}{6}$&$0$&$-\tfrac{2}{3}$&$-\tfrac{5}{6}$&$-\tfrac{1}{3}$&$\tfrac{2}{3}$&$0$&$-\tfrac{2}{3}$&$\tfrac{19}{18}$&$-1$\\
$\bar{n}_{4}$&$5$&$1$&$0$&$0$&$0$&$-\tfrac{1}{6}$&${\color{red}-\tfrac{4}{3}\normalcolor}$&$-\tfrac{1}{2}$&$\left(\mathbf{1},\,\mathbf{1},\,\mathbf{1},\,\mathbf{1}\right)$&$0$&$\tfrac{1}{6}$&$0$&$-\tfrac{2}{3}$&$-\tfrac{5}{6}$&$\tfrac{2}{3}$&$\tfrac{2}{3}$&$0$&$\tfrac{1}{3}$&$\tfrac{19}{18}$&$1$\\
\hline
$y_{2}$&$5$&$0$&$1$&$1$&$0$&$-\tfrac{1}{6}$&$-\tfrac{1}{3}$&${\color{red}\tfrac{1}{2}\normalcolor}$&$\left(\mathbf{1},\,\mathbf{2},\,\mathbf{1},\,\mathbf{2}\right)$&$0$&$-\tfrac{1}{3}$&$0$&$0$&$0$&$\tfrac{1}{6}$&$0$&$\tfrac{1}{2}$&$0$&$\tfrac{13}{18}$&$0$\\
$m_{3}$&$5$&$0$&$1$&$1$&$0$&$-\tfrac{1}{6}$&$-\tfrac{1}{3}$&${\color{red}\tfrac{1}{2}\normalcolor}$&$\left(\mathbf{1},\,\mathbf{2},\,\mathbf{1},\,\mathbf{1}\right)$&$0$&$-\tfrac{1}{3}$&$0$&$0$&$0$&$\tfrac{1}{6}$&$1$&$-\tfrac{1}{2}$&$0$&$-\tfrac{17}{18}$&$0$\\
$m_{4}$&$5$&$0$&$1$&$1$&$0$&$-\tfrac{1}{6}$&$-\tfrac{1}{3}$&${\color{red}\tfrac{1}{2}\normalcolor}$&$\left(\mathbf{1},\,\mathbf{2},\,\mathbf{1},\,\mathbf{1}\right)$&$0$&$-\tfrac{1}{3}$&$0$&$0$&$0$&$\tfrac{1}{6}$&$-1$&$-\tfrac{1}{2}$&$0$&$-\tfrac{29}{18}$&$0$\\
\hline
$y_{1}$&$5$&$0$&$1$&$0$&$0$&$-\tfrac{1}{6}$&$-\tfrac{1}{3}$&${\color{red}\tfrac{1}{2}\normalcolor}$&$\left(\mathbf{1},\,\mathbf{2},\,\mathbf{1},\,\mathbf{2}\right)$&$0$&$-\tfrac{1}{3}$&$0$&$0$&$0$&$\tfrac{1}{6}$&$0$&$\tfrac{1}{2}$&$0$&$\tfrac{13}{18}$&$0$\\
$m_{1}$&$5$&$0$&$1$&$0$&$0$&$-\tfrac{1}{6}$&$-\tfrac{1}{3}$&${\color{red}\tfrac{1}{2}\normalcolor}$&$\left(\mathbf{1},\,\mathbf{2},\,\mathbf{1},\,\mathbf{1}\right)$&$0$&$-\tfrac{1}{3}$&$0$&$0$&$0$&$\tfrac{1}{6}$&$1$&$-\tfrac{1}{2}$&$0$&$-\tfrac{17}{18}$&$0$\\
$m_{2}$&$5$&$0$&$1$&$0$&$0$&$-\tfrac{1}{6}$&$-\tfrac{1}{3}$&${\color{red}\tfrac{1}{2}\normalcolor}$&$\left(\mathbf{1},\,\mathbf{2},\,\mathbf{1},\,\mathbf{1}\right)$&$0$&$-\tfrac{1}{3}$&$0$&$0$&$0$&$\tfrac{1}{6}$&$-1$&$-\tfrac{1}{2}$&$0$&$-\tfrac{29}{18}$&$0$\\
\hline
$\bar{d}_{1}$&$5$&$0$&$0$&$1$&$0$&$-\tfrac{1}{6}$&$-\tfrac{1}{3}$&$-\tfrac{1}{2}$&$\left(\mathbf{\bar{3}},\,\mathbf{1},\,\mathbf{1},\,\mathbf{1}\right)$&$-\tfrac{1}{3}$&$\tfrac{1}{6}$&$0$&$0$&$-\tfrac{3}{2}$&$-\tfrac{1}{3}$&$0$&$0$&$0$&$-\tfrac{5}{18}$&$\tfrac{1}{3}$\\
$\bar{e}_{1}$&$5$&$0$&$0$&$1$&$0$&$-\tfrac{1}{6}$&$-\tfrac{1}{3}$&$-\tfrac{1}{2}$&$\left(\mathbf{1},\,\mathbf{1},\,\mathbf{1},\,\mathbf{1}\right)$&$-1$&$\tfrac{1}{6}$&$0$&$0$&$\tfrac{1}{2}$&$-\tfrac{1}{3}$&$0$&$0$&$0$&$\tfrac{7}{18}$&$-1$\\
$\bar{u}_{1}$&$5$&$0$&$0$&$1$&$0$&$-\tfrac{1}{6}$&$-\tfrac{1}{3}$&$-\tfrac{1}{2}$&$\left(\mathbf{\bar{3}},\,\mathbf{1},\,\mathbf{1},\,\mathbf{1}\right)$&$\tfrac{2}{3}$&$\tfrac{1}{6}$&$0$&$0$&$\tfrac{1}{2}$&$-\tfrac{1}{3}$&$0$&$0$&$0$&$\tfrac{7}{18}$&$\tfrac{1}{3}$\\
$l_{1}$&$5$&$0$&$0$&$1$&$0$&$-\tfrac{1}{6}$&$-\tfrac{1}{3}$&$-\tfrac{1}{2}$&$\left(\mathbf{1},\,\mathbf{2},\,\mathbf{1},\,\mathbf{1}\right)$&$\tfrac{1}{2}$&$\tfrac{1}{6}$&$0$&$0$&$-\tfrac{3}{2}$&$-\tfrac{1}{3}$&$0$&$0$&$0$&$-\tfrac{5}{18}$&$1$\\
$q_{1}$&$5$&$0$&$0$&$1$&$0$&$-\tfrac{1}{6}$&$-\tfrac{1}{3}$&$-\tfrac{1}{2}$&$\left(\mathbf{3},\,\mathbf{2},\,\mathbf{1},\,\mathbf{1}\right)$&$-\tfrac{1}{6}$&$\tfrac{1}{6}$&$0$&$0$&$\tfrac{1}{2}$&$-\tfrac{1}{3}$&$0$&$0$&$0$&$\tfrac{7}{18}$&$-\tfrac{1}{3}$\\
$\bar{n}_{1}$&$5$&$0$&$0$&$1$&$0$&$-\tfrac{1}{6}$&$-\tfrac{1}{3}$&$-\tfrac{1}{2}$&$\left(\mathbf{1},\,\mathbf{1},\,\mathbf{1},\,\mathbf{1}\right)$&$0$&$\tfrac{1}{6}$&$0$&$0$&$\tfrac{5}{2}$&$-\tfrac{1}{3}$&$0$&$0$&$0$&$\tfrac{19}{18}$&$-1$\\
$s^0_{12}$&$5$&$0$&$0$&$1$&$0$&$\tfrac{5}{6}$&$-\tfrac{1}{3}$&$-\tfrac{1}{2}$&$\left(\mathbf{1},\,\mathbf{1},\,\mathbf{1},\,\mathbf{1}\right)$&$0$&$\tfrac{2}{3}$&$-\tfrac{1}{2}$&$\tfrac{1}{2}$&$0$&$-\tfrac{1}{3}$&$0$&$0$&$0$&$-\tfrac{5}{18}$&$0$\\
$s^0_{11}$&$5$&$0$&$0$&$1$&$0$&$\tfrac{5}{6}$&$-\tfrac{1}{3}$&$-\tfrac{1}{2}$&$\left(\mathbf{1},\,\mathbf{1},\,\mathbf{1},\,\mathbf{1}\right)$&$0$&$\tfrac{2}{3}$&$\tfrac{1}{2}$&$-\tfrac{1}{2}$&$0$&$-\tfrac{1}{3}$&$0$&$0$&$0$&$\tfrac{25}{18}$&$0$\\
$s^0_{14}$&$5$&$0$&$0$&$1$&$0$&$\tfrac{11}{6}$&$-\tfrac{1}{3}$&$-\tfrac{1}{2}$&$\left(\mathbf{1},\,\mathbf{1},\,\mathbf{1},\,\mathbf{1}\right)$&$0$&$-\tfrac{1}{3}$&$-\tfrac{1}{2}$&$-\tfrac{1}{2}$&$0$&$-\tfrac{1}{3}$&$0$&$0$&$0$&$\tfrac{13}{18}$&$0$\\
$s^0_{13}$&$5$&$0$&$0$&$1$&$0$&$\tfrac{11}{6}$&$-\tfrac{1}{3}$&$-\tfrac{1}{2}$&$\left(\mathbf{1},\,\mathbf{1},\,\mathbf{1},\,\mathbf{1}\right)$&$0$&$-\tfrac{1}{3}$&$\tfrac{1}{2}$&$\tfrac{1}{2}$&$0$&$-\tfrac{1}{3}$&$0$&$0$&$0$&$-\tfrac{17}{18}$&$0$\\
$s^0_{10}$&$5$&$0$&$0$&$1$&$0$&$-\tfrac{1}{6}$&$\tfrac{2}{3}$&$-\tfrac{1}{2}$&$\left(\mathbf{1},\,\mathbf{1},\,\mathbf{1},\,\mathbf{1}\right)$&$0$&$-\tfrac{1}{3}$&$-\tfrac{1}{2}$&$-\tfrac{1}{2}$&$0$&$-\tfrac{1}{3}$&$0$&$0$&$0$&$\tfrac{13}{18}$&$0$\\
$s^0_{9}$&$5$&$0$&$0$&$1$&$0$&$-\tfrac{1}{6}$&$\tfrac{2}{3}$&$-\tfrac{1}{2}$&$\left(\mathbf{1},\,\mathbf{1},\,\mathbf{1},\,\mathbf{1}\right)$&$0$&$-\tfrac{1}{3}$&$\tfrac{1}{2}$&$\tfrac{1}{2}$&$0$&$-\tfrac{1}{3}$&$0$&$0$&$0$&$-\tfrac{17}{18}$&$0$\\
\hline
$\bar{d}_{2}$&$5$&$0$&$0$&$0$&$0$&$-\tfrac{1}{6}$&$-\tfrac{1}{3}$&$-\tfrac{1}{2}$&$\left(\mathbf{\bar{3}},\,\mathbf{1},\,\mathbf{1},\,\mathbf{1}\right)$&$-\tfrac{1}{3}$&$\tfrac{1}{6}$&$0$&$0$&$-\tfrac{3}{2}$&$-\tfrac{1}{3}$&$0$&$0$&$0$&$-\tfrac{5}{18}$&$\tfrac{1}{3}$\\
$\bar{e}_{2}$&$5$&$0$&$0$&$0$&$0$&$-\tfrac{1}{6}$&$-\tfrac{1}{3}$&$-\tfrac{1}{2}$&$\left(\mathbf{1},\,\mathbf{1},\,\mathbf{1},\,\mathbf{1}\right)$&$-1$&$\tfrac{1}{6}$&$0$&$0$&$\tfrac{1}{2}$&$-\tfrac{1}{3}$&$0$&$0$&$0$&$\tfrac{7}{18}$&$-1$\\
$\bar{u}_{2}$&$5$&$0$&$0$&$0$&$0$&$-\tfrac{1}{6}$&$-\tfrac{1}{3}$&$-\tfrac{1}{2}$&$\left(\mathbf{\bar{3}},\,\mathbf{1},\,\mathbf{1},\,\mathbf{1}\right)$&$\tfrac{2}{3}$&$\tfrac{1}{6}$&$0$&$0$&$\tfrac{1}{2}$&$-\tfrac{1}{3}$&$0$&$0$&$0$&$\tfrac{7}{18}$&$\tfrac{1}{3}$\\
$l_{2}$&$5$&$0$&$0$&$0$&$0$&$-\tfrac{1}{6}$&$-\tfrac{1}{3}$&$-\tfrac{1}{2}$&$\left(\mathbf{1},\,\mathbf{2},\,\mathbf{1},\,\mathbf{1}\right)$&$\tfrac{1}{2}$&$\tfrac{1}{6}$&$0$&$0$&$-\tfrac{3}{2}$&$-\tfrac{1}{3}$&$0$&$0$&$0$&$-\tfrac{5}{18}$&$1$\\
$q_{2}$&$5$&$0$&$0$&$0$&$0$&$-\tfrac{1}{6}$&$-\tfrac{1}{3}$&$-\tfrac{1}{2}$&$\left(\mathbf{3},\,\mathbf{2},\,\mathbf{1},\,\mathbf{1}\right)$&$-\tfrac{1}{6}$&$\tfrac{1}{6}$&$0$&$0$&$\tfrac{1}{2}$&$-\tfrac{1}{3}$&$0$&$0$&$0$&$\tfrac{7}{18}$&$-\tfrac{1}{3}$\\
$\bar{n}_{2}$&$5$&$0$&$0$&$0$&$0$&$-\tfrac{1}{6}$&$-\tfrac{1}{3}$&$-\tfrac{1}{2}$&$\left(\mathbf{1},\,\mathbf{1},\,\mathbf{1},\,\mathbf{1}\right)$&$0$&$\tfrac{1}{6}$&$0$&$0$&$\tfrac{5}{2}$&$-\tfrac{1}{3}$&$0$&$0$&$0$&$\tfrac{19}{18}$&$-1$\\
$s^0_{6}$&$5$&$0$&$0$&$0$&$0$&$\tfrac{5}{6}$&$-\tfrac{1}{3}$&$-\tfrac{1}{2}$&$\left(\mathbf{1},\,\mathbf{1},\,\mathbf{1},\,\mathbf{1}\right)$&$0$&$\tfrac{2}{3}$&$-\tfrac{1}{2}$&$\tfrac{1}{2}$&$0$&$-\tfrac{1}{3}$&$0$&$0$&$0$&$-\tfrac{5}{18}$&$0$\\
$s^0_{5}$&$5$&$0$&$0$&$0$&$0$&$\tfrac{5}{6}$&$-\tfrac{1}{3}$&$-\tfrac{1}{2}$&$\left(\mathbf{1},\,\mathbf{1},\,\mathbf{1},\,\mathbf{1}\right)$&$0$&$\tfrac{2}{3}$&$\tfrac{1}{2}$&$-\tfrac{1}{2}$&$0$&$-\tfrac{1}{3}$&$0$&$0$&$0$&$\tfrac{25}{18}$&$0$\\
$s^0_{8}$&$5$&$0$&$0$&$0$&$0$&$\tfrac{11}{6}$&$-\tfrac{1}{3}$&$-\tfrac{1}{2}$&$\left(\mathbf{1},\,\mathbf{1},\,\mathbf{1},\,\mathbf{1}\right)$&$0$&$-\tfrac{1}{3}$&$-\tfrac{1}{2}$&$-\tfrac{1}{2}$&$0$&$-\tfrac{1}{3}$&$0$&$0$&$0$&$\tfrac{13}{18}$&$0$\\
$s^0_{7}$&$5$&$0$&$0$&$0$&$0$&$\tfrac{11}{6}$&$-\tfrac{1}{3}$&$-\tfrac{1}{2}$&$\left(\mathbf{1},\,\mathbf{1},\,\mathbf{1},\,\mathbf{1}\right)$&$0$&$-\tfrac{1}{3}$&$\tfrac{1}{2}$&$\tfrac{1}{2}$&$0$&$-\tfrac{1}{3}$&$0$&$0$&$0$&$-\tfrac{17}{18}$&$0$\\
$s^0_{4}$&$5$&$0$&$0$&$0$&$0$&$-\tfrac{1}{6}$&$\tfrac{2}{3}$&$-\tfrac{1}{2}$&$\left(\mathbf{1},\,\mathbf{1},\,\mathbf{1},\,\mathbf{1}\right)$&$0$&$-\tfrac{1}{3}$&$-\tfrac{1}{2}$&$-\tfrac{1}{2}$&$0$&$-\tfrac{1}{3}$&$0$&$0$&$0$&$\tfrac{13}{18}$&$0$\\
$s^0_{3}$&$5$&$0$&$0$&$0$&$0$&$-\tfrac{1}{6}$&$\tfrac{2}{3}$&$-\tfrac{1}{2}$&$\left(\mathbf{1},\,\mathbf{1},\,\mathbf{1},\,\mathbf{1}\right)$&$0$&$-\tfrac{1}{3}$&$\tfrac{1}{2}$&$\tfrac{1}{2}$&$0$&$-\tfrac{1}{3}$&$0$&$0$&$0$&$-\tfrac{17}{18}$&$0$\\
\hline
$\bar{\nu}_{2}$&$5$&$1$&$1$&$1$&$0$&$-\tfrac{1}{6}$&${\color{red}-\tfrac{4}{3}\normalcolor}$&$-\tfrac{1}{2}$&$\left(\mathbf{\bar{3}},\,\mathbf{1},\,\mathbf{1},\,\mathbf{1}\right)$&$\tfrac{1}{6}$&$-\tfrac{1}{3}$&$0$&$-\tfrac{1}{3}$&$\tfrac{1}{3}$&$\tfrac{1}{6}$&$\tfrac{1}{3}$&$\tfrac{1}{2}$&$-\tfrac{1}{3}$&$\tfrac{25}{18}$&$-\tfrac{2}{3}$\\
$s^-_{8}$&$5$&$1$&$1$&$1$&$0$&$-\tfrac{1}{6}$&${\color{red}-\tfrac{4}{3}\normalcolor}$&${\color{red}\tfrac{1}{2}\normalcolor}$&$\left(\mathbf{1},\,\mathbf{1},\,\mathbf{1},\,\mathbf{1}\right)$&$\tfrac{1}{2}$&$-\tfrac{1}{3}$&$0$&$\tfrac{2}{3}$&$-\tfrac{2}{3}$&$\tfrac{1}{6}$&$\tfrac{1}{3}$&$\tfrac{1}{2}$&$-\tfrac{1}{3}$&$-\tfrac{11}{18}$&$0$\\
$s^+_{7}$&$5$&$1$&$1$&$1$&$0$&$-\tfrac{1}{6}$&${\color{red}\tfrac{2}{3}\normalcolor}$&${\color{red}\tfrac{1}{2}\normalcolor}$&$\left(\mathbf{1},\,\mathbf{1},\,\mathbf{1},\,\mathbf{1}\right)$&$-\tfrac{1}{2}$&$\tfrac{1}{6}$&$-\tfrac{1}{2}$&$\tfrac{1}{6}$&$-\tfrac{1}{6}$&$\tfrac{1}{6}$&$\tfrac{1}{3}$&$-\tfrac{1}{2}$&$\tfrac{2}{3}$&$-\tfrac{17}{18}$&$1$\\
$m_{8}$&$5$&$1$&$1$&$1$&$0$&$-\tfrac{1}{6}$&${\color{red}\tfrac{2}{3}\normalcolor}$&$-\tfrac{1}{2}$&$\left(\mathbf{1},\,\mathbf{2},\,\mathbf{1},\,\mathbf{1}\right)$&$0$&$\tfrac{1}{6}$&$\tfrac{1}{2}$&$\tfrac{1}{6}$&$\tfrac{5}{6}$&$\tfrac{1}{6}$&$\tfrac{1}{3}$&$\tfrac{1}{2}$&$-\tfrac{1}{3}$&$\tfrac{19}{18}$&$-1$\\
$s^-_{7}$&$5$&$1$&$1$&$1$&$0$&$-\tfrac{1}{6}$&$-\tfrac{1}{3}$&$-\tfrac{1}{2}$&$\left(\mathbf{1},\,\mathbf{1},\,\mathbf{1},\,\mathbf{1}\right)$&$\tfrac{1}{2}$&$\tfrac{2}{3}$&$0$&$-\tfrac{1}{3}$&$-\tfrac{2}{3}$&$\tfrac{1}{6}$&$\tfrac{1}{3}$&$\tfrac{1}{2}$&$-\tfrac{1}{3}$&$\tfrac{31}{18}$&$0$\\
$s^+_{8}$&$5$&$1$&$1$&$1$&$0$&$\tfrac{5}{6}$&${\color{red}-\tfrac{4}{3}\normalcolor}$&$-\tfrac{1}{2}$&$\left(\mathbf{1},\,\mathbf{1},\,\mathbf{1},\,\mathbf{1}\right)$&$-\tfrac{1}{2}$&$\tfrac{1}{6}$&$-\tfrac{1}{2}$&$\tfrac{1}{6}$&$-\tfrac{1}{6}$&$\tfrac{1}{6}$&$\tfrac{1}{3}$&$\tfrac{1}{2}$&$-\tfrac{1}{3}$&$\tfrac{13}{18}$&$-1$\\
\hline
$\bar{\nu}_{1}$&$5$&$1$&$1$&$0$&$0$&$-\tfrac{1}{6}$&${\color{red}-\tfrac{4}{3}\normalcolor}$&$-\tfrac{1}{2}$&$\left(\mathbf{\bar{3}},\,\mathbf{1},\,\mathbf{1},\,\mathbf{1}\right)$&$\tfrac{1}{6}$&$-\tfrac{1}{3}$&$0$&$-\tfrac{1}{3}$&$\tfrac{1}{3}$&$\tfrac{1}{6}$&$\tfrac{1}{3}$&$\tfrac{1}{2}$&$-\tfrac{1}{3}$&$\tfrac{25}{18}$&$-\tfrac{2}{3}$\\
$s^-_{6}$&$5$&$1$&$1$&$0$&$0$&$-\tfrac{1}{6}$&${\color{red}-\tfrac{4}{3}\normalcolor}$&${\color{red}\tfrac{1}{2}\normalcolor}$&$\left(\mathbf{1},\,\mathbf{1},\,\mathbf{1},\,\mathbf{1}\right)$&$\tfrac{1}{2}$&$-\tfrac{1}{3}$&$0$&$\tfrac{2}{3}$&$-\tfrac{2}{3}$&$\tfrac{1}{6}$&$\tfrac{1}{3}$&$\tfrac{1}{2}$&$-\tfrac{1}{3}$&$-\tfrac{11}{18}$&$0$\\
$s^+_{5}$&$5$&$1$&$1$&$0$&$0$&$-\tfrac{1}{6}$&${\color{red}\tfrac{2}{3}\normalcolor}$&${\color{red}\tfrac{1}{2}\normalcolor}$&$\left(\mathbf{1},\,\mathbf{1},\,\mathbf{1},\,\mathbf{1}\right)$&$-\tfrac{1}{2}$&$\tfrac{1}{6}$&$-\tfrac{1}{2}$&$\tfrac{1}{6}$&$-\tfrac{1}{6}$&$\tfrac{1}{6}$&$\tfrac{1}{3}$&$-\tfrac{1}{2}$&$\tfrac{2}{3}$&$-\tfrac{17}{18}$&$1$\\
$m_{7}$&$5$&$1$&$1$&$0$&$0$&$-\tfrac{1}{6}$&${\color{red}\tfrac{2}{3}\normalcolor}$&$-\tfrac{1}{2}$&$\left(\mathbf{1},\,\mathbf{2},\,\mathbf{1},\,\mathbf{1}\right)$&$0$&$\tfrac{1}{6}$&$\tfrac{1}{2}$&$\tfrac{1}{6}$&$\tfrac{5}{6}$&$\tfrac{1}{6}$&$\tfrac{1}{3}$&$\tfrac{1}{2}$&$-\tfrac{1}{3}$&$\tfrac{19}{18}$&$-1$\\
$s^-_{5}$&$5$&$1$&$1$&$0$&$0$&$-\tfrac{1}{6}$&$-\tfrac{1}{3}$&$-\tfrac{1}{2}$&$\left(\mathbf{1},\,\mathbf{1},\,\mathbf{1},\,\mathbf{1}\right)$&$\tfrac{1}{2}$&$\tfrac{2}{3}$&$0$&$-\tfrac{1}{3}$&$-\tfrac{2}{3}$&$\tfrac{1}{6}$&$\tfrac{1}{3}$&$\tfrac{1}{2}$&$-\tfrac{1}{3}$&$\tfrac{31}{18}$&$0$\\
$s^+_{6}$&$5$&$1$&$1$&$0$&$0$&$\tfrac{5}{6}$&${\color{red}-\tfrac{4}{3}\normalcolor}$&$-\tfrac{1}{2}$&$\left(\mathbf{1},\,\mathbf{1},\,\mathbf{1},\,\mathbf{1}\right)$&$-\tfrac{1}{2}$&$\tfrac{1}{6}$&$-\tfrac{1}{2}$&$\tfrac{1}{6}$&$-\tfrac{1}{6}$&$\tfrac{1}{6}$&$\tfrac{1}{3}$&$\tfrac{1}{2}$&$-\tfrac{1}{3}$&$\tfrac{13}{18}$&$-1$\\
\hline
$f_{3}$&$5$&$1$&$0$&$1$&$0$&$-\tfrac{1}{6}$&${\color{red}\tfrac{2}{3}\normalcolor}$&$-\tfrac{1}{2}$&$\left(\mathbf{1},\,\mathbf{1},\,\mathbf{4},\,\mathbf{1}\right)$&$0$&$\tfrac{1}{6}$&$0$&$-\tfrac{1}{3}$&$\tfrac{5}{6}$&$\tfrac{1}{6}$&$\tfrac{1}{3}$&$-\tfrac{1}{2}$&$\tfrac{1}{6}$&$\tfrac{1}{18}$&$0$\\
$\bar{f}_{3}$&$5$&$1$&$0$&$1$&$0$&$-\tfrac{1}{6}$&${\color{red}\tfrac{2}{3}\normalcolor}$&$-\tfrac{1}{2}$&$\left(\mathbf{1},\,\mathbf{1},\,\mathbf{\bar{4}},\,\mathbf{1}\right)$&$0$&$\tfrac{1}{6}$&$0$&$-\tfrac{1}{3}$&$\tfrac{5}{6}$&$\tfrac{1}{6}$&$\tfrac{1}{3}$&$\tfrac{1}{2}$&$\tfrac{1}{6}$&$\tfrac{37}{18}$&$0$\\
$\bar{\eta}_{2}$&$5$&$1$&$0$&$1$&$0$&$\tfrac{5}{6}$&$-\tfrac{1}{3}$&$-\tfrac{1}{2}$&$\left(\mathbf{1},\,\mathbf{1},\,\mathbf{1},\,\mathbf{2}\right)$&$0$&$\tfrac{1}{6}$&$0$&$-\tfrac{1}{3}$&$\tfrac{5}{6}$&$-\tfrac{1}{3}$&$\tfrac{1}{3}$&$0$&$-\tfrac{1}{3}$&$\tfrac{19}{18}$&$-1$\\
$n_{4}$&$5$&$1$&$0$&$1$&$0$&$\tfrac{5}{6}$&$-\tfrac{1}{3}$&$-\tfrac{1}{2}$&$\left(\mathbf{1},\,\mathbf{1},\,\mathbf{1},\,\mathbf{1}\right)$&$0$&$\tfrac{1}{6}$&$0$&$-\tfrac{1}{3}$&$\tfrac{5}{6}$&$-\tfrac{1}{3}$&$-\tfrac{2}{3}$&$0$&$\tfrac{2}{3}$&$\tfrac{19}{18}$&$1$\\
$\bar{n}_{7}$&$5$&$1$&$0$&$1$&$0$&$\tfrac{5}{6}$&${\color{red}\tfrac{2}{3}\normalcolor}$&$-\tfrac{1}{2}$&$\left(\mathbf{1},\,\mathbf{1},\,\mathbf{1},\,\mathbf{1}\right)$&$0$&$\tfrac{1}{6}$&$0$&$-\tfrac{1}{3}$&$\tfrac{5}{6}$&$\tfrac{2}{3}$&$-\tfrac{2}{3}$&$0$&$-\tfrac{1}{3}$&$\tfrac{7}{18}$&$-1$\\
\hline
$f_{2}$&$5$&$1$&$0$&$0$&$0$&$-\tfrac{1}{6}$&${\color{red}\tfrac{2}{3}\normalcolor}$&$-\tfrac{1}{2}$&$\left(\mathbf{1},\,\mathbf{1},\,\mathbf{4},\,\mathbf{1}\right)$&$0$&$\tfrac{1}{6}$&$0$&$-\tfrac{1}{3}$&$\tfrac{5}{6}$&$\tfrac{1}{6}$&$\tfrac{1}{3}$&$-\tfrac{1}{2}$&$\tfrac{1}{6}$&$\tfrac{1}{18}$&$0$\\
$\bar{f}_{2}$&$5$&$1$&$0$&$0$&$0$&$-\tfrac{1}{6}$&${\color{red}\tfrac{2}{3}\normalcolor}$&$-\tfrac{1}{2}$&$\left(\mathbf{1},\,\mathbf{1},\,\mathbf{\bar{4}},\,\mathbf{1}\right)$&$0$&$\tfrac{1}{6}$&$0$&$-\tfrac{1}{3}$&$\tfrac{5}{6}$&$\tfrac{1}{6}$&$\tfrac{1}{3}$&$\tfrac{1}{2}$&$\tfrac{1}{6}$&$\tfrac{37}{18}$&$0$\\
$\bar{\eta}_{1}$&$5$&$1$&$0$&$0$&$0$&$\tfrac{5}{6}$&$-\tfrac{1}{3}$&$-\tfrac{1}{2}$&$\left(\mathbf{1},\,\mathbf{1},\,\mathbf{1},\,\mathbf{2}\right)$&$0$&$\tfrac{1}{6}$&$0$&$-\tfrac{1}{3}$&$\tfrac{5}{6}$&$-\tfrac{1}{3}$&$\tfrac{1}{3}$&$0$&$-\tfrac{1}{3}$&$\tfrac{19}{18}$&$-1$\\
$n_{3}$&$5$&$1$&$0$&$0$&$0$&$\tfrac{5}{6}$&$-\tfrac{1}{3}$&$-\tfrac{1}{2}$&$\left(\mathbf{1},\,\mathbf{1},\,\mathbf{1},\,\mathbf{1}\right)$&$0$&$\tfrac{1}{6}$&$0$&$-\tfrac{1}{3}$&$\tfrac{5}{6}$&$-\tfrac{1}{3}$&$-\tfrac{2}{3}$&$0$&$\tfrac{2}{3}$&$\tfrac{19}{18}$&$1$\\
$\bar{n}_{6}$&$5$&$1$&$0$&$0$&$0$&$\tfrac{5}{6}$&${\color{red}\tfrac{2}{3}\normalcolor}$&$-\tfrac{1}{2}$&$\left(\mathbf{1},\,\mathbf{1},\,\mathbf{1},\,\mathbf{1}\right)$&$0$&$\tfrac{1}{6}$&$0$&$-\tfrac{1}{3}$&$\tfrac{5}{6}$&$\tfrac{2}{3}$&$-\tfrac{2}{3}$&$0$&$-\tfrac{1}{3}$&$\tfrac{7}{18}$&$-1$
\end{longtable} \endgroup \end{onehalfspacing} \normalsize

\bibliography{z6_2}

\providecommand{\href}[2]{#2}\begingroup\raggedright\begin{thebibliography}{10}

\bibitem{Nilles:2013lda}
H.~P. Nilles, S.~{Ramos-Sanchez}, M.~Ratz, and P.~K. Vaudrevange, ``{A note on
  discrete $R$ symmetries in $\mathbb{Z}$$_{6}$-II orbifolds with Wilson
  lines},'' \href{http://dx.doi.org/10.1016/j.physletb.2013.09.041}{{\em
  Phys.Lett.} {\bfseries B726} (2013) 876--881},
\href{http://arxiv.org/abs/1308.3435}{{\ttfamily arXiv:1308.3435 [hep-th]}}.
%%CITATION = ARXIV:1308.3435;%%.

\bibitem{Bizet:2013wha}
N.~G. Cabo~Bizet, T.~Kobayashi, D.~K. Mayorga~{Pena}, S.~L. Parameswaran,
  M.~Schmitz, {\em et~al.}, ``{Discrete R-symmetries and Anomaly Universality
  in Heterotic Orbifolds},''
  \href{http://dx.doi.org/10.1007/JHEP02(2014)098}{{\em JHEP} {\bfseries 1402}
  (2014) 098},
\href{http://arxiv.org/abs/1308.5669}{{\ttfamily arXiv:1308.5669 [hep-th]}}.
%%CITATION = ARXIV:1308.5669;%%.

\bibitem{Lebedev:2006kn}
O.~Lebedev, H.~P. Nilles, S.~Raby, S.~Ramos-Sanchez, M.~Ratz, {\em et~al.},
  ``{A Mini-landscape of exact MSSM spectra in heterotic orbifolds},''
  \href{http://dx.doi.org/10.1016/j.physletb.2006.12.012}{{\em Phys.Lett.}
  {\bfseries B645} (2007) 88--94},
\href{http://arxiv.org/abs/hep-th/0611095}{{\ttfamily arXiv:hep-th/0611095
  [hep-th]}}.
%%CITATION = HEP-TH/0611095;%%.

\bibitem{Lebedev:2007hv}
O.~Lebedev, H.~P. Nilles, S.~Raby, S.~Ramos-Sanchez, M.~Ratz, {\em et~al.},
  ``{The Heterotic Road to the MSSM with R parity},''
  \href{http://dx.doi.org/10.1103/PhysRevD.77.046013}{{\em Phys.Rev.}
  {\bfseries D77} (2008) 046013},
\href{http://arxiv.org/abs/0708.2691}{{\ttfamily arXiv:0708.2691 [hep-th]}}.
%%CITATION = ARXIV:0708.2691;%%.

\bibitem{Lebedev:2008un}
O.~Lebedev, H.~P. Nilles, S.~Ramos-Sanchez, M.~Ratz, and P.~K. Vaudrevange,
  ``{Heterotic mini-landscape. (II). Completing the search for MSSM vacua in a
  Z(6) orbifold},''
  \href{http://dx.doi.org/10.1016/j.physletb.2008.08.054}{{\em Phys.Lett.}
  {\bfseries B668} (2008) 331--335},
\href{http://arxiv.org/abs/0807.4384}{{\ttfamily arXiv:0807.4384 [hep-th]}}.
%%CITATION = ARXIV:0807.4384;%%.

\bibitem{Kappl:2011vi}
R.~Kappl, M.~Ratz, and C.~Staudt, ``{The Hilbert basis method for D-flat
  directions and the superpotential},''
  \href{http://dx.doi.org/10.1007/JHEP10(2011)027}{{\em JHEP} {\bfseries 1110}
  (2011) 027},
\href{http://arxiv.org/abs/1108.2154}{{\ttfamily arXiv:1108.2154 [hep-th]}}.
%%CITATION = ARXIV:1108.2154;%%.

\bibitem{Kappl:2010yu}
R.~Kappl, B.~Petersen, S.~Raby, M.~Ratz, R.~Schieren, {\em et~al.},
  ``{String-Derived MSSM Vacua with Residual R Symmetries},''
  \href{http://dx.doi.org/10.1016/j.nuclphysb.2011.01.032}{{\em Nucl.Phys.}
  {\bfseries B847} (2011) 325--349},
\href{http://arxiv.org/abs/1012.4574}{{\ttfamily arXiv:1012.4574 [hep-th]}}.
%%CITATION = ARXIV:1012.4574;%%.

\bibitem{Antusch:2014poa}
S.~Antusch, I.~de~Medeiros~Varzielas, V.~Maurer, C.~Sluka, and M.~Spinrath,
  ``{Towards predictive flavour models in SUSY SU(5) GUTs with doublet-triplet
  splitting},''
\href{http://arxiv.org/abs/1405.6962}{{\ttfamily arXiv:1405.6962 [hep-ph]}}.
%%CITATION = ARXIV:1405.6962;%%.

\bibitem{normaliz:2014}
W.~Bruns, B.~Ichim, T.~{R\"omer}, and C.~{S\"oger}, ``Normaliz. {Algorithms}
  for rational cones and affine monoids.'' Available from
  http://www.math.uos.de/normaliz.

\bibitem{4ti2}
4ti2 team, ``4ti2---a software package for algebraic, geometric and
  combinatorial problems on linear spaces.'' {A}vailable at www.4ti2.de.

\bibitem{2012arXiv1206.1916B}
W.~{Bruns}, B.~{Ichim}, and C.~{S{\"o}ger}, ``{The power of pyramid
  decomposition in Normaliz},'' {\em ArXiv e-prints} (June, 2012) ,
  \href{http://arxiv.org/abs/1206.1916}{{\ttfamily arXiv:1206.1916 [math.CO]}}.

\bibitem{Chen:2012st}
M.-C. Chen, J.~Huang, J.-M. O'Bryan, A.~M. Wijangco, and F.~Yu,
  ``{Compatibility of $\theta_{13}$ and the Type I Seesaw Model with $A_4$
  Symmetry},'' \href{http://dx.doi.org/10.1007/JHEP02(2013)021}{{\em JHEP}
  {\bfseries 1302} (2013) 021},
\href{http://arxiv.org/abs/1210.6982}{{\ttfamily arXiv:1210.6982 [hep-ph]}}.
%%CITATION = ARXIV:1210.6982;%%.

\bibitem{Buchmuller:2006ik}
W.~Buchmuller, K.~Hamaguchi, O.~Lebedev, and M.~Ratz, ``{Supersymmetric
  Standard Model from the Heterotic String (II)},''
  \href{http://dx.doi.org/10.1016/j.nuclphysb.2007.06.028}{{\em Nucl.Phys.}
  {\bfseries B785} (2007) 149--209},
\href{http://arxiv.org/abs/hep-th/0606187}{{\ttfamily arXiv:hep-th/0606187
  [hep-th]}}.
%%CITATION = HEP-TH/0606187;%%.

\bibitem{Bruns20101098}
W.~Bruns and B.~Ichim, ``Normaliz: Algorithms for affine monoids and rational
  cones,''
  \href{http://dx.doi.org/http://dx.doi.org/10.1016/j.jalgebra.2010.01.031}{{\em
  Journal of Algebra} {\bfseries 324} no.~5, (2010) 1098 -- 1113}.
  Computational Algebra.

\bibitem{Kappl:2008ie}
R.~Kappl, H.~P. Nilles, S.~Ramos-Sanchez, M.~Ratz, K.~Schmidt-Hoberg, {\em
  et~al.}, ``{Large hierarchies from approximate R symmetries},''
  \href{http://dx.doi.org/10.1103/PhysRevLett.102.121602}{{\em Phys.Rev.Lett.}
  {\bfseries 102} (2009) 121602},
\href{http://arxiv.org/abs/0812.2120}{{\ttfamily arXiv:0812.2120 [hep-th]}}.
%%CITATION = ARXIV:0812.2120;%%.

\bibitem{Brummer:2010fr}
F.~Brummer, R.~Kappl, M.~Ratz, and K.~Schmidt-Hoberg, ``{Approximate
  R-symmetries and the mu term},''
  \href{http://dx.doi.org/10.1007/JHEP04(2010)006}{{\em JHEP} {\bfseries 1004}
  (2010) 006},
\href{http://arxiv.org/abs/1003.0084}{{\ttfamily arXiv:1003.0084 [hep-th]}}.
%%CITATION = ARXIV:1003.0084;%%.

\bibitem{Ko:2007dz}
P.~Ko, T.~Kobayashi, J.-h. Park, and S.~Raby, ``{String-derived D(4) flavor
  symmetry and phenomenological implications},''
  \href{http://dx.doi.org/10.1103/PhysRevD.76.059901,
  10.1103/PhysRevD.76.035005, 10.1103/PhysRevD.76.035005
  10.1103/PhysRevD.76.059901}{{\em Phys.Rev.} {\bfseries D76} (2007) 035005},
\href{http://arxiv.org/abs/0704.2807}{{\ttfamily arXiv:0704.2807 [hep-ph]}}.
%%CITATION = ARXIV:0704.2807;%%.

\bibitem{Kobayashi:2006wq}
T.~Kobayashi, H.~P. Nilles, F.~Ploger, S.~Raby, and M.~Ratz, ``{Stringy origin
  of non-Abelian discrete flavor symmetries},''
  \href{http://dx.doi.org/10.1016/j.nuclphysb.2007.01.018}{{\em Nucl.Phys.}
  {\bfseries B768} (2007) 135--156},
\href{http://arxiv.org/abs/hep-ph/0611020}{{\ttfamily arXiv:hep-ph/0611020
  [hep-ph]}}.
%%CITATION = HEP-PH/0611020;%%.

\bibitem{Bizet:2013gf}
N.~G. Cabo~Bizet, T.~Kobayashi, D.~K. Mayorga~Pena, S.~L. Parameswaran,
  M.~Schmitz, {\em et~al.}, ``{R-charge Conservation and More in Factorizable
  and Non-Factorizable Orbifolds},''
  \href{http://dx.doi.org/10.1007/JHEP05(2013)076}{{\em JHEP} {\bfseries 1305}
  (2013) 076},
\href{http://arxiv.org/abs/1301.2322}{{\ttfamily arXiv:1301.2322 [hep-th]}}.
%%CITATION = ARXIV:1301.2322;%%.

\bibitem{Nilles:2011aj}
H.~P. Nilles, S.~Ramos-Sanchez, P.~K. Vaudrevange, and A.~Wingerter, ``{The
  Orbifolder: A Tool to study the Low Energy Effective Theory of Heterotic
  Orbifolds},'' \href{http://dx.doi.org/10.1016/j.cpc.2012.01.026}{{\em
  Comput.Phys.Commun.} {\bfseries 183} (2012) 1363--1380},
\href{http://arxiv.org/abs/1110.5229}{{\ttfamily arXiv:1110.5229 [hep-th]}}.
%%CITATION = ARXIV:1110.5229;%%.

\bibitem{Maniatis:2006jd}
M.~Maniatis, A.~von Manteuffel, and O.~Nachtmann, ``{Determining the global
  minimum of Higgs potentials via Groebner bases: Applied to the NMSSM},''
  \href{http://dx.doi.org/10.1140/epjc/s10052-006-0186-2}{{\em Eur.Phys.J.}
  {\bfseries C49} (2007) 1067--1076},
\href{http://arxiv.org/abs/hep-ph/0608314}{{\ttfamily arXiv:hep-ph/0608314
  [hep-ph]}}.
%%CITATION = HEP-PH/0608314;%%.

\bibitem{Gray:2006gn}
J.~Gray, Y.-H. He, and A.~Lukas, ``{Algorithmic Algebraic Geometry and Flux
  Vacua},'' \href{http://dx.doi.org/10.1088/1126-6708/2006/09/031}{{\em JHEP}
  {\bfseries 0609} (2006) 031},
\href{http://arxiv.org/abs/hep-th/0606122}{{\ttfamily arXiv:hep-th/0606122
  [hep-th]}}.
%%CITATION = HEP-TH/0606122;%%.

\bibitem{Gray:2007yq}
J.~Gray, Y.-H. He, A.~Ilderton, and A.~Lukas, ``{A New Method for Finding Vacua
  in String Phenomenology},''
  \href{http://dx.doi.org/10.1088/1126-6708/2007/07/023}{{\em JHEP} {\bfseries
  0707} (2007) 023},
\href{http://arxiv.org/abs/hep-th/0703249}{{\ttfamily arXiv:hep-th/0703249
  [HEP-TH]}}.
%%CITATION = HEP-TH/0703249;%%.

\bibitem{dgps}
W.~Decker, G.-M. Greuel, G.~Pfister, and H.~Sch\"onemann, ``{\sc Singular}
  {3-1-6} --- {A} computer algebra system for polynomial computations.''
  {http://www.singular.uni-kl.de}, 2012.

\bibitem{Lee:2010gv}
H.~M. Lee, S.~Raby, M.~Ratz, G.~G. Ross, R.~Schieren, {\em et~al.}, ``{A unique
  $Z_4^R$ symmetry for the MSSM},''
  \href{http://dx.doi.org/10.1016/j.physletb.2010.10.038}{{\em Phys.Lett.}
  {\bfseries B694} (2011) 491--495},
\href{http://arxiv.org/abs/1009.0905}{{\ttfamily arXiv:1009.0905 [hep-ph]}}.
%%CITATION = ARXIV:1009.0905;%%.

\bibitem{Dreiner:2013ala}
H.~K. Dreiner, T.~Opferkuch, and C.~Luhn, ``{Froggatt-Nielsen models with a
  residual $\mathbb{Z}^R_4$ symmetry},''
  \href{http://dx.doi.org/10.1103/PhysRevD.88.115005}{{\em Phys.Rev.}
  {\bfseries D88} no.~11, (2013) 115005},
\href{http://arxiv.org/abs/1308.0332}{{\ttfamily arXiv:1308.0332 [hep-ph]}}.
%%CITATION = ARXIV:1308.0332;%%.

\bibitem{Chen:2014gua}
M.-C. Chen, M.~Ratz, and V.~Takhistov, ``{$R$ parity violation from discrete
  $R$ symmetries},''
\href{http://arxiv.org/abs/1410.3474}{{\ttfamily arXiv:1410.3474 [hep-ph]}}.
%%CITATION = ARXIV:1410.3474;%%.

\bibitem{Camargo-Molina:2013qva}
J.~Camargo-Molina, B.~O'Leary, W.~Porod, and F.~Staub, ``{$\mathbf{Vevacious}$:
  A Tool For Finding The Global Minima Of One-Loop Effective Potentials With
  Many Scalars},'' \href{http://dx.doi.org/10.1140/epjc/s10052-013-2588-2}{{\em
  Eur.Phys.J.} {\bfseries C73} (2013) 2588},
\href{http://arxiv.org/abs/1307.1477}{{\ttfamily arXiv:1307.1477 [hep-ph]}}.
%%CITATION = ARXIV:1307.1477;%%.

\bibitem{Cicoli:2013cha}
M.~Cicoli, D.~Klevers, S.~Krippendorf, C.~Mayrhofer, F.~Quevedo, {\em et~al.},
  ``{Explicit de Sitter Flux Vacua for Global String Models with Chiral
  Matter},'' \href{http://dx.doi.org/10.1007/JHEP05(2014)001}{{\em JHEP}
  {\bfseries 1405} (2014) 001},
\href{http://arxiv.org/abs/1312.0014}{{\ttfamily arXiv:1312.0014 [hep-th]}}.
%%CITATION = ARXIV:1312.0014;%%.

\bibitem{Buccella:1982nx}
F.~Buccella, J.~Derendinger, S.~Ferrara, and C.~A. Savoy, ``{Patterns of
  Symmetry Breaking in Supersymmetric Gauge Theories},''
\href{http://dx.doi.org/10.1016/0370-2693(82)90521-4}{{\em Phys.Lett.}
  {\bfseries B115} (1982) 375}.
%%CITATION = PHLTA,B115,375;%%.

\bibitem{Cleaver:1997jb}
G.~Cleaver, M.~Cvetic, J.~R. Espinosa, L.~L. Everett, and P.~Langacker,
  ``{Classification of flat directions in perturbative heterotic superstring
  vacua with anomalous U(1)},''
  \href{http://dx.doi.org/10.1016/S0550-3213(98)00277-6}{{\em Nucl.Phys.}
  {\bfseries B525} (1998) 3--26},
\href{http://arxiv.org/abs/hep-th/9711178}{{\ttfamily arXiv:hep-th/9711178
  [hep-th]}}.
%%CITATION = HEP-TH/9711178;%%.

\bibitem{Hosteins:2009xk}
P.~Hosteins, R.~Kappl, M.~Ratz, and K.~Schmidt-Hoberg, ``{Gauge-top
  unification},'' \href{http://dx.doi.org/10.1088/1126-6708/2009/07/029}{{\em
  JHEP} {\bfseries 0907} (2009) 029},
\href{http://arxiv.org/abs/0905.3323}{{\ttfamily arXiv:0905.3323 [hep-ph]}}.
%%CITATION = ARXIV:0905.3323;%%.

\bibitem{Kappl:2010sg}
R.~Kappl, ``{Quark mass hierarchies in heterotic orbifold GUTs},''
  \href{http://dx.doi.org/10.1007/JHEP04(2011)019}{{\em JHEP} {\bfseries 1104}
  (2011) 019},
\href{http://arxiv.org/abs/1012.4368}{{\ttfamily arXiv:1012.4368 [hep-ph]}}.
%%CITATION = ARXIV:1012.4368;%%.

\bibitem{Buchmuller:2007zd}
W.~Buchmuller, K.~Hamaguchi, O.~Lebedev, S.~Ramos-Sanchez, and M.~Ratz,
  ``{Seesaw neutrinos from the heterotic string},''
  \href{http://dx.doi.org/10.1103/PhysRevLett.99.021601}{{\em Phys.Rev.Lett.}
  {\bfseries 99} (2007) 021601},
\href{http://arxiv.org/abs/hep-ph/0703078}{{\ttfamily arXiv:hep-ph/0703078
  [HEP-PH]}}.
%%CITATION = HEP-PH/0703078;%%.

\bibitem{Heeck:2012fw}
J.~Heeck, ``{Seesaw parametrization for n right-handed neutrinos},''
  \href{http://dx.doi.org/10.1103/PhysRevD.86.093023}{{\em Phys.Rev.}
  {\bfseries D86} (2012) 093023},
\href{http://arxiv.org/abs/1207.5521}{{\ttfamily arXiv:1207.5521 [hep-ph]}}.
%%CITATION = ARXIV:1207.5521;%%.

\end{thebibliography}\endgroup
\bibliographystyle{utphys}

\end{document}